\documentclass{aa}
\usepackage{graphicx}
\usepackage{subfig}
\usepackage{txfonts}
\usepackage{longtable}

\usepackage[normalem]{ulem}
\def\ha{H$\alpha$}
\newcommand{\nii}{[N\,{\sc{ii}}]}

\usepackage{color}

\begin{document}

   \title{GLACE survey: OSIRIS/GTC tuneable imaging of the galaxy cluster ZwCl 0024.0+1652}

   \subtitle{II. The mass--metallicity relationship and the effect of the environment}

      \author{Bernab\'e Cedr\'es
          \inst{1,2}
    \and
      Simon B. De Daniloff
          \inst{1}
   \and
     \'Angel Bongiovanni 
          \inst{1,2}
    \and
      Miguel S\'anchez-Portal
          \inst{1,2}
    \and
    Miguel Cervi\~no \inst{3}
    \and
    Ricardo P\'erez-Mart\'inez
          \inst{4,2}
    \and
    Ana Mar{\'i}a P\'erez-Garc{\'i}a
         \inst{3,2}
    \and
    Jordi Cepa
        \inst{5,6,2}
    \and
    Maritza A. Lara-L\'opez
        \inst{7}
    \and
    Mauro Gonz\'alez-Otero
        \inst{5,6,2}
    \and
    Manuel Castillo-Fraile
        \inst{1}
    \and
    Jos\'e Ignacio Gonz\'alez-Serrano
        \inst{8,2}
    \and
    Castalia Alenka Negrete
        \inst{9}
    \and
    Camen P. Padilla-Torres
        \inst{5,6,2,10}
    \and
    Irene Pintos-Castro
        \inst{11}
    \and
    Mirjana Povi\'c
        \inst{12,13,14}
    \and
    Emilio Alfaro
        \inst{13}
    \and
    Zeleke Beyoro-Amado
        \inst{12,15,16}
    \and
    Irene Cruz-Gonz\'alez
        \inst{8}
    \and
    Jos\'e A. de Diego
        \inst{9}
    \and
    Roc{\'i}o Navarro Mart{\'i}nez
        \inst{2}
    \and
    Brisa Mancillas
        \inst{1}
    \and
    M\'onica I. Rodr{\'i}guez
        \inst{1}
    \and
    Iv\'an Valtchanov
        \inst{17}
          }
   \institute{Institut de Radioastronomie Millim\'etrique (IRAM), Av. Divina Pastora 7, N\'ucleo Central 18012, Granada, Spain\\
              \email{bcedres@iram.es}
        \and
            Asociaci\'on Astrof{\'i}sica para la Promoci\'on de la Investigaci\'on, Instrumentaci\'on y su Desarrollo, ASPID, 38205 La Laguna, Tenerife, Spain
        \and
            Centro de Astrobiolog{\'i}a (CSIC/INTA), 28692 ESAC Campus, Villanueva de la Cañada, Madrid, Spain
        \and
            ISDEFE for European Space Astronomy Centre (ESAC)/ESA, P.O. Box 78, E-28690 Vi\-lla\-nue\-va de la Ca\~nada, Madrid, Spain
        \and 
            Instituto de Astrof{\'i}sica de Canarias (IAC), 38200 La Laguna, Tenerife, Spain
         \and
             Departamento de Astrof{\'i}sica, Universidad de La Laguna (ULL), 38205 La Laguna, Tenerife, Spain
        \and
            Departamento de F{\'i}sica de la Tierra y Astrof{\'i}sica, Instituto de F{\'i}sica de Part{\'i}culas y del Cosmos, IPARCOS. Universidad Complutense de Madrid (UCM), 28040 Madrid, Spain
       \and
            Instituto de F{\'i}sica de Cantabria (CSIC-Universidad de Cantabria), 39005, Santander, Spain
        \and
            Instituto de Astronom{\'i}a, Universidad Nacional Aut\'onoma de
            M\'exico, Apdo. Postal 70-264, 04510 Ciudad de M\'exico, M\'exico
        \and
            Fundaci\'on Galileo Galilei-INAF Rambla Jos\'e Ana Fern\'andez P\'erez, 7, 38712 Bre\~na Baja, Tenerife, Spain
        \and
            Centro de Estudios de {\'i}sica del Cosmos de Arag\'on (CEFCA),
            Plaza San Juan 1, 44001 Teruel, Spain
        \and
            Space Science and Geospatial Institute (SSGI), Entoto
            Observatory and Research Center (EORC), Astronomy and Astrophysics Research Division, PO Box 33679, Addis Abbaba, Ethiopia
        \and
            Instituto de Astrof{\'i}sica de Andaluc{\'i}a (CSIC), 18080 Granada, Spain
        \and
            Physics Department, Mbarara University of Science and Technology (MUST), Mbarara, Uganda
        \and
            Addis Ababa University (AAU), P.O.Box 1176, Addis Ababa, Ethiopia
        \and
            Physics Department, Kotebe Metropolitan University (KMU) P.O.Box 31248, Addis Ababa, Ethiopia
        \and
            Telespazio Vega UK for ESA, European Space Astronomy Centre, Operations Departmen, E-28691, Villanueva de la Ca\~nada, Spain
       } 
   \date{}

% \abstract{}{}{}{}{} 
% 5 {} token are mandatory
 
  \abstract
   {}
   {In this paper, we revisit the data for the galaxy cluster ZwCl 0024.0+1652  provided by the GLACE survey and study the mass--metallicity function and its relationship with the environment.}
   {Here we describe an alternative way to reduce the data from OSIRIS tunable filters. This method gives us better uncertainties in the fluxes of the emission lines and the derived quantities. We present an updated catalogue of cluster galaxies with emission in \ha\ and \nii\  $\lambda\lambda$6548,6583. We also discuss the biases of these new fluxes and describe the way in which we calculated the mass--metallicity relationship and its uncertainties.}
   {We generated a new catalogue of 84 emission-line galaxies  with reliable fluxes in \nii\  and \ha\  lines from a list of 174 galaxies. We find a relationship between the clustercentric radius and the density of galaxies. We derived the mass--metallicity relationship for ZwCl 0024.0+1652 and compared it with clusters and field galaxies from the literature. We find a difference in the mass--metallicity relationship when compared to more massive clusters, with the latter showing on average higher values of abundance. This could be an effect of the quenching of the star formation, which seems to be more prevalent in low-mass galaxies in more massive clusters. We find little to no difference between ZwCl 0024.0+1652 galaxies and field galaxies located at the same redshift.}
   {}


   \titlerunning{GLACE survey: Mass--Metallicity relationship}
   \maketitle
%
%-------------------------------------------------------------------

\section{Introduction}

The evolution of the star formation process in cluster galaxies is still an open field. The pioneering works of \cite{gisler78} and \cite{dressler1985} established that emission-line galaxies (ELGs) are more numerous in the field than in clusters. Other works have delved deeper into the issue: for example \cite{balogh1997}  found that cluster galaxies are likely to have less significant star formation processes than field galaxies, finding that the star formation rate (SFR) in cluster galaxies is lower when compared with field galaxies; \cite{cucciati2010} and \cite{cantale2016} reported that the fraction of blue galaxies is significantly lower in groups than in field galaxies; \cite{allen2016} found that the mass-normalised SFR for the highest-mass galaxies in the field was larger than for galaxies in clusters;  \cite{old2020} found that the star-forming main sequence for galaxies in clusters is lower compared to that of the field galaxies, and this effect was more significant for lower-mass galaxies; and \cite{vaug2020} found that the average \ha-to-continuum-size ratio of ELGs is smaller in cluster galaxies than in field galaxies with the same stellar mass.

The reasons for these differences may be associated with the quenching of star formation in cluster galaxies. \cite{boselli2005} list the possible causes in their review: starvation (removal of the outer galaxy halo that feeds the star formation, preventing further infall of gas into the disc (\citealt{larson1980}); ram-pressure stripping (removal of the galactic gas by moving at high velocities in dense and hot intergalactic medium (\citealt{gunn1972}); or harassment (interactions of the galaxy with the potential of the cluster as a whole (\citealt{moore1996}), among others.

Metallicities have been found to be affected by this quenching (see \citealt{darvish2015}). There is a tight correlation between the abundance of galaxies and their stellar mass: the more massive the galaxy, the higher the metallicity. This is called the mass--metallicity relationship (hereafter, MZR). As proposed by \cite{laralopez2010}, this relationship may be an indicator of a more profound link between metallicity, stellar mass, and SFR, which has been referred to as the `fundamental plane'.

It is assumed that pristine gas falls into galaxies and is then turned into stars and released at the end of the life of the massive stars. Some of this gas, enriched with metals, may escape the gravity well of its parent galaxy. However, in more massive galaxies with larger gravity wells, this gas will have a lesser tendency to escape when compared with galaxies of lower masses. Therefore, in general, in the absence of interactions with other galaxies, this may imply that galaxies with larger stellar masses will be more metallic. Following this criterion, the MZR may be influenced by the environment: galaxies located within the denser areas of a cluster could potentially be subject to a greater number of interactions than field galaxies or galaxies in the outskirts of the same cluster. These interactions may strip galaxies of their less metallic gas by some of the processes mentioned above, and these galaxies will therefore have a higher metallicity content than those in the field or in less dense environments (see e.g. \citealt{Maier2019}, and references therein).
The MZR has been described in many different contexts; for example, in field galaxies in the local Universe (\citealt{tremonti2004}, \citealt{lee2006} or \citealt{andrews2013}, among others), in field galaxies at higher redshifts (i.e. \citealt{calabro2017} for $0.1<z<0.9$ or \citealt{hayashi2009} for up to $z\sim2$), and in galaxies in clusters (see e.g. \citealt{sobral2016}, \citealt{ciocan2020}, \citealt{Maier2019}, \citealt{maritza2022} or \citealt{perez2023}, among others).

Here, we present a revisiting of the cluster ZwCl\,0024.0+1652 (hereafter Cl0024) using improved reduction techniques.
The paper is organised as follows. In Section 2, we describe the GLACE survey, explain the inverse convolution method, and present the galaxy cluster Cl0024. In Section 3, we discuss the way we parameterised the environment. Section 4 is devoted to the MZR for Cl0024, where we also present   a comparison with other clusters and field galaxies from the literature. In Section 5 we discuss our main results and present the conclusions of this work.
Throughout this paper, we assume a standard $\Lambda$-cold dark matter cosmology with $\Omega_\Lambda = 0.7$, $\Omega_m = 0.3$, and $\mathrm{H}_0 = 70 \,\mathrm{km}\,  \mathrm{s}^{-1}\, \mathrm{Mpc}^{-1}$.
%--------------------------------------------------------------------
\section{The GLACE survey: Revisiting the pseudospectra}
%RPM:
\label{sec:GLACE}
The GLACE (GaLAxy Cluster Evolution Survey) is a narrow-band survey of ELGs and AGNs in a sample of galaxy clusters located at different redshifts. The aim of GLACE is to study the variations in galaxy properties as a function of environment. The main objectives and methods of this survey are described in detail in \cite{glace}. Briefly, GLACE was developed as the cluster counterpart of the OTELO survey (\citealt{otelo}), observing clusters at $z \sim 0.4, 0.63,$ and $0.86$, and mapping several rest-frame optical emission lines, such as \ha, H$\beta$, \nii, [O\,{\sc{ii}}]$\lambda3727,$ and [O\,{\sc{iii}}]$\lambda5007$. 

We selected  Cl0024 from the lower redshift observations; it has a precise redshift of $z=0.395$. We obtained data through tunable filter tomography, employing the instrument OSIRIS on the 10.4 meter GTC (Gran Telescopio Canarias) at Roque de los Muchachos Observatory. The whole process of obtaining and reducing the data is described in \cite{glace}.
Briefly, the data were obtained through the technique of TF tomography (\citealt{jordi2013}), where a series of tuned images were obtained, sampling the \nii\ and \ha\ lines. The result is a pseudospectrum for each ELG. This pseudospectrum is a convolution of the real spectrum of the ELG with the response function of the OSIRIS instrument. A total of 174 ELGs were detected from the cluster. From there, fluxes from the \ha\, and \nii\, lines were derived, as well as redshifts. AGNs were detected and marked as such (see \citealt{glace} for details). 

\subsection{The inverse convolution method}
The method to derive the fluxes and their uncertainty for the 174 ELGs is detailed in Eqs 10 and 11 of \cite{glace}; in Eq. 11, the uncertainty is obtained using a first-order approximation of the variance in uncertainty propagation theory. However, such first-order approximations lead to 
larger, unrealistic variance estimates, especially when ratios of quantities are used and the nominal variances of the quantities involved have an associated relative standard deviation of greater than 10\% (as is the case in Eq. 10 of  \cite{glace}). Worse yet, for large relative standard deviations, not only is the uncertainty overestimated, but the nominal value of the inferred quantity is also biased if ratios or logarithm operations are involved \cite[see][for a case of study for different inference types]{CVG03}. 

Fortunately, since this first iteration with the data, a new method to obtain the fluxes from the pseudospectra has been developed. This latter is based on a Monte Carlo sampling of the multi-parametric probability distribution function (PDF) of the  intensity of the continuum fluxes and the shape of the principal lines in the pseudo-spectrum produced by the tunable filters. The multi-parametric PDFs are also used to obtain the associated PDFs of line ratios and related quantities, such as metallicities, and this approach has been employed in several works related to the OTELO Survey (i.e. \citealt{bongio2020}, \citealt{jakub2020}, \citealt{beli2021}, and \citealt{rocio2021}).

This method, named `inverse convolution', is described in detail in Appendix A of \cite{jakub2020}.
Briefly, a model spectrum as a rest-frame spectrum
is simulated by Gaussian profiles of the \nii\ and \ha\ lines defined by their amplitude, common line width for three lines, and a constant continuum level. The model is then convolved with the spectral response of the OSIRIS instrument and then compared with the observational data through a likelihood function. This is repeated a total of $10^5$ times. After this process, we obtain a PDF shape for each parameter, as well as all the required additional quantities. This allows us to obtain more reliable values for the \ha\ and \nii\ fluxes, which will lead to more informed estimates of the metallicity. The inverse convolutions can be used to derive the redshift, the continuum, the fluxes of the lines (\ha, \nii$\lambda6548$ and \nii$\lambda6583$ in this case), and the full width at half maximum (FWHM) of the lines. The uncertainties can be directly obtained from the PDF. In our case, we use the 68\% confidence interval around the mode of the PDF for each parameter. 

\begin{figure}[]
   \centering
   \includegraphics[width=\hsize]{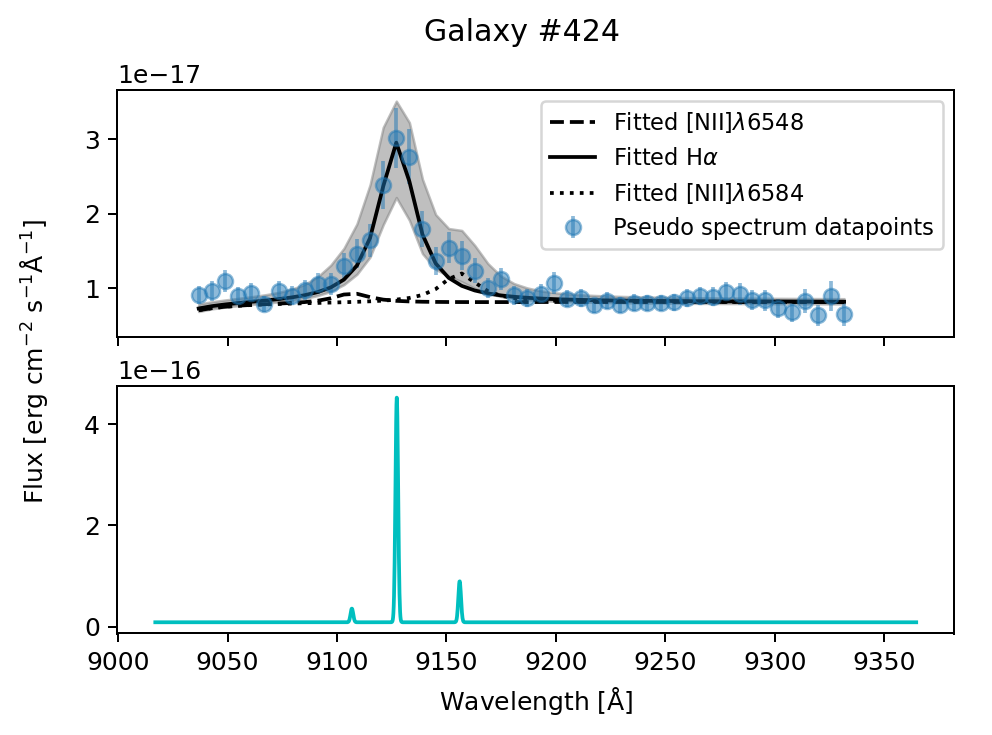}
      \caption{Pseudospectrum with fitting components (upper panel) and deconvolved spectrum (lower panel) for the tier 1 galaxy {\tt id:424}. The filled grey area represents the 68\% confidence interval.}
         \label{esp1494}
\end{figure}

This new way of obtaining the fluxes from the emission lines allows us to revisit the data from Cl0024 and study its properties in greater detail. The results can be seen in Fig. \ref{esp1494}, where present the pseudospectrum of the galaxy {\tt id:424} (upper panel) and the deconvolved spectrum (lower panel) as well as the 68\% confidence interval of the fit. 

After all the galaxies from \cite{glace} were reprocessed by the inverse convolution method, a classification of the fitted spectra ---based on careful visual inspection--- was carried out, taking into account the quality of the deconvolution as well as the uncertainties. In this new classification, the galaxies were sorted into tiers. Tiers 1 to 3 are well-deconvolved galaxies, with galaxies from tier 1 having the smallest uncertainties and those from tier 3 galaxies having the largest. We also made sure that all fluxes, redshifts, and derived quantities present PDFs that can be approximated as Gaussians. This means that the mode, median, and mean of the PDFs are very close, and so we can use any of them interchangeably. From the total of 174 galaxies by \cite{glace}, we derived reliable fluxes for 84 of them. For the remaining galaxies,  in
47 cases we were unable to separate the \nii\, lines from the \ha\, line, 16 cases presented a complex \ha\, line made up of several superimposing components, and in 27 cases the algorithm did not converge. Of the cases with a composite \ha\, line and those for which we were unable to separate \ha\, and \nii\, lines,  75\% and 30\% are classified as AGNs by \cite{glace} respectively. Table \ref{numgal} contains a summary of the classification for all the galaxies. In Table \ref{Tablagorda}, we present the results of the inverse convolution for the 84 galaxies from tiers 1 to 3, indicating the \ha\ and \nii$\lambda6584$ fluxes as well as the metallicity. Hereafter, unless otherwise stated, we only analyse and discuss galaxies from tiers 1 to 3.

\begin{table}[]
    \centering
    \caption{Number of galaxies in the sample segregated by quality.}
    {\renewcommand{\arraystretch}{1.3}%
    \begin{tabular}{lc}
    \hline
    \hline
    Quality tier & Number of galaxies\\
    \hline
     Tier 1    &  14\\
     Tier 2    &  38\\
     Tier 3    &  32\\
     \hline
     Total deconvolved & 84\\
     \hline
     No deblending & 47\\
     Composite \ha\ line & 16\\
     Unable to deconvolve & 27\\
     \hline
     Total galaxies & 174 \\
     \hline
    \end{tabular}}
    \tablefoot{Only galaxies with tiers 1 to 3 are considered to have reliable fluxes for the \ha\ and \nii\ lines.}
    \label{numgal}
\end{table}

\subsection{Galaxy cluster ZwCl 0024.0+1652}

Cl0024 is a galaxy cluster located at mid-redshift ($z\sim0.4$); it has been described as having a rather complex structure that does not appear at visible wavelengths (\citealt{czoske2002}), and is low in X-ray flux and central velocity dispersion (\citealt{johnson2016}).
Cl0024 is one of the original clusters of the pioneering paper by \cite{bo1978}, where the so-called Butcher--Oemler effect was first described. \cite{zeleke2021} reported that the core is almost totally made up of early-type galaxies with no emission lines up to $\sim500$\,kpc, with increasing numbers of late-type galaxies in the outskirts. \cite{glace} suggest that this seems to indicate a quenching of star formation and AGN activity.

The cluster has been divided into at least two substructures by several authors (i.e. \citealt{czoske2002}, \citealt{moran2007}, \citealt{glace}): one that constitutes the main cluster and centred at $z=0.395$ (thereafter Structure A) and a second group of infalling galaxies at $z=0.381$ (thereafter Structure B). \cite{czoske2002} suggest that these two structures are colliding along the line of sight. Moreover, \cite{czoske2002} and \cite{glace} propose the existence of a third interacting structure at about $z\sim0.42$. 

\begin{figure}[]
   \centering
   \includegraphics[width=\hsize]{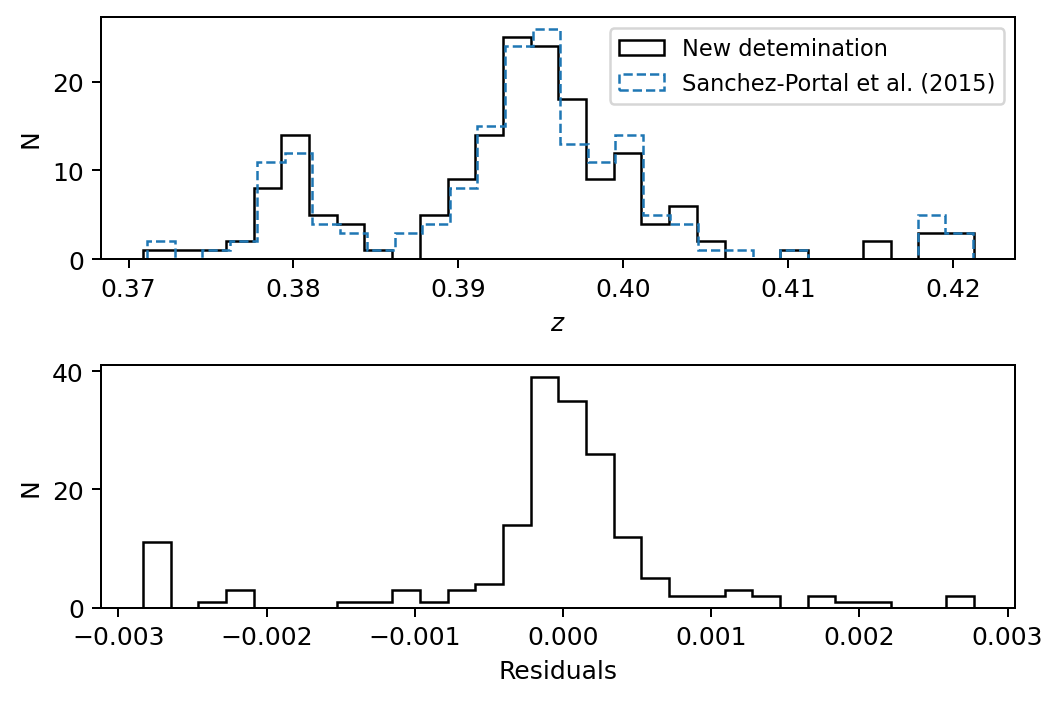}
      \caption{Redshift distribution for the galaxies presented in this study of Cl0024. The upper panel shows the distribution of redshifts derived from the new determination (solid line) and the determination from \cite{glace} (dashed line). The lower panel shows the distribution of the residuals between the old and the new redshift determination. This figure includes all the galaxies (tiers 1 to 3, and `no deblending' and `composite \ha\, lines').}
         \label{fig_redshift}
\end{figure}

In the upper panel of Fig.\ref{fig_redshift}, we show the distribution of redshifts derived from the inverse convolution method (solid black line) and the redshifts from \cite{glace} (blue dashed line). The two structures are clearly identified in the histogram. Moreover, the proposed third structure can be seen at $z\sim0.42$. The lower panel shows the residuals between the new and the old determination. The coincidence between both methods is very good, with the majority of the galaxies presenting differences of less than 0.001 in redshift. 

\begin{figure}[]
   \centering
   \includegraphics[width=\hsize]{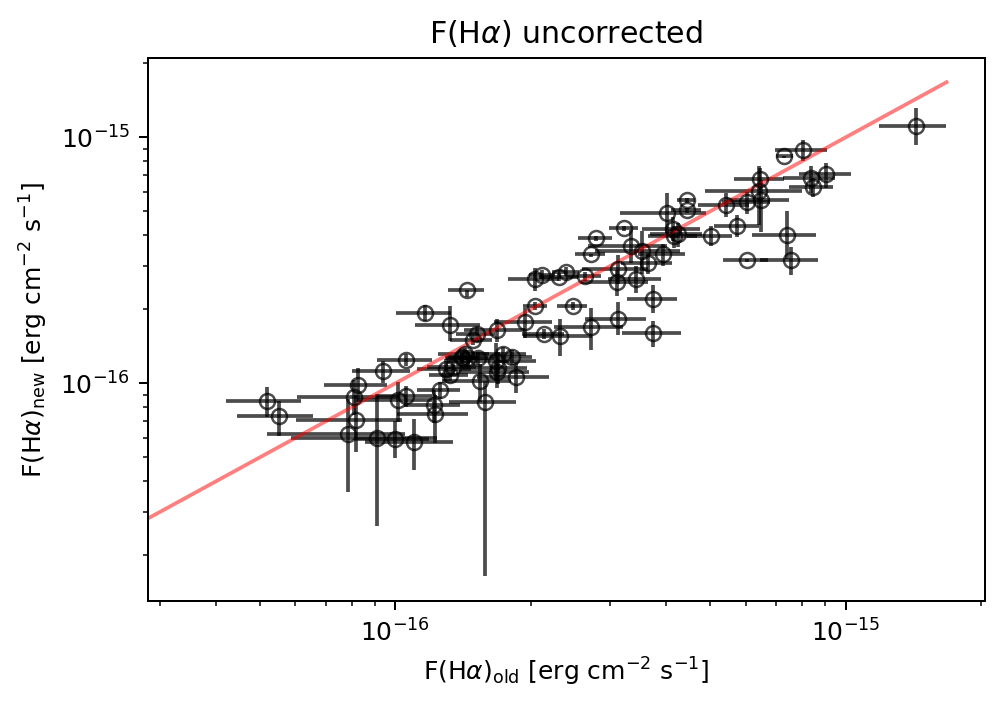}
      \caption{Comparison between the obtained values for the \ha\ emission line uncorrected for extinction from \cite{glace} and the new one from this work. The red continuous line indicates a 1:1 relation.}
         \label{relacha}
\end{figure}

Figure \ref{relacha} shows the \ha\ flux derived from the inverse convolution method versus the \ha\ flux derived from \cite{glace}. There is a very good correlation between the results of the two methods (as expected). Nevertheless, it seems that the determination by \cite{glace} has marginally higher values when compared with those from the inverse convolution method.

\begin{figure}[]
   \centering
   \includegraphics[width=\hsize]{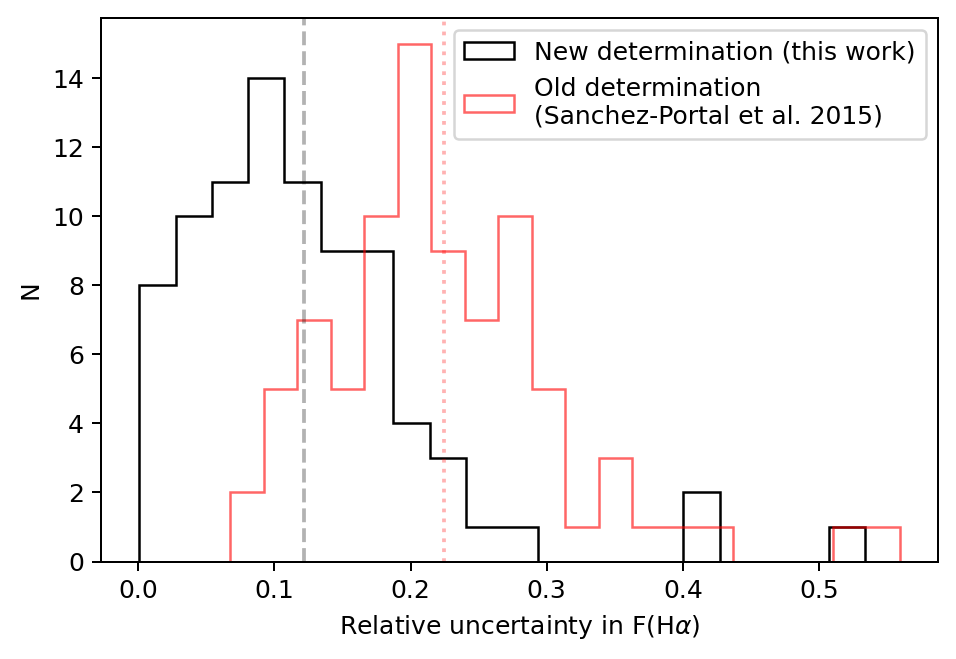}
      \caption{Histogram of the relative uncertainty in the flux determination in the \ha\ line. The red histogram is the error from \cite{glace}. The red dotted line is the median value of the relative error. The black histogram is the relative error from this work. The grey dashed line represents the median value.}
         \label{histha}
\end{figure}

On the other hand, the real strength of the inverse convolution method is shown in Fig. \ref{histha} where it is compared to the distribution of the relative uncertainties obtained by \cite{glace} and the new ones, defined as the 68\% confidence interval. The median value of the distribution of relative uncertainties in the \ha\ flux from the \cite{glace} determination is above 20\%, while with the inverse convolution method is just over 10\%, with many galaxies having uncertainties in the \ha\ line at 5\% and below. This difference allows us to calculate all the derived quantities (such as the abundance) with enough precision to to make robust conclusions.

\section{Parameterisation of the environment}
\label{sec:environment}

We characterise the local environment of our sample by applying a modified version of the method first described in \cite{dressler1980}. We calculated the projected surface density of the galaxies in our data set, estimated as the source density in the area encircled between the object and the tenth nearest galaxy above a certain magnitude threshold  ($\Sigma_{10}$) according to the following Eq.:
\begin{equation}
    \Sigma_{10}=\frac{10+1}{\pi R_{10}^2},
    \label{densidad}
\end{equation}
where $R_{10}$ is the projected distance in megaparsecs from the considered galaxy to the tenth closest object. In this case, we take objects brighter than I $\sim$ 21.5 measured in GLACE ancillary data (\citealt{glace}), where the completeness of the spectroscopic member catalogue is above 70\% along the full area covered by the observations (Perez-Martinez et al. in prep). We also take into account the two main structures described in section \ref{sec:GLACE}.

\begin{figure}[]
    \centering
    \includegraphics[width=\hsize]{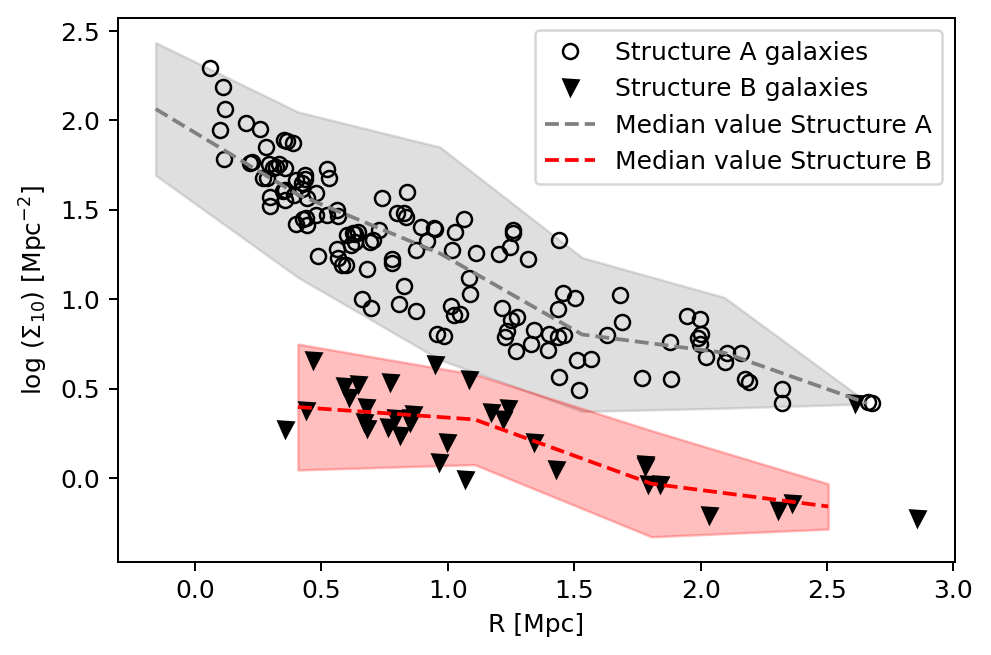}    \caption{Logarithm of $\Sigma_{10}$ as a function of the clustercentric radius for all the galaxies from \cite{glace}. Open circles and downward-pointing filled triangles are the galaxies from Structure A and B, respectively. The grey and red dashed lines show the median value for galaxies in Structures A and B, respectively. The filled grey and red areas represent the median absolute deviation for Structures A and B, respectively.}
    \label{sigmavsr}
\end{figure}

Figure \ref{sigmavsr} shows the logarithm of the $\Sigma_{10}$ as a function of the clustercentric radius for all galaxies from \cite{glace}. We take the centre of the cluster to be that suggested in \cite{glace}. The median values for the Structure A and B galaxies are indicated by the grey dashed line and the red dashed line, respectively. A tight correlation exists, with galaxies with a lower clustercentric radius having a larger value in $\log(\Sigma_{10})$. This is not surprising because, we expect to find a denser environment
for galaxies in the inner parts of the cluster. The galaxies from the infall Structure show, in general, lower values in density when compared with galaxies in the main structure and can be clearly identified in Fig. \ref{sigmavsr}.  We note that the clustercentric radius used in this case originates from the centre of Structure A, and so, in principle, we would expect a greater scattering of galaxies in Structure B. However, the figure shows a tight correlation for those galaxies. This seems to indicate that the centres of the two structures must be relatively closely projected onto the sky, as suggested by \cite{czoske2002}.

\section{The mass--metallicity relationship}\label{sec:mzr}
To determine the abundance of the ELGs of Cl0024, and keeping in mind that we are restricted to the \ha\, and \nii\, lines, we employed the recipe given by \cite{pp04} using the ratio $\ensuremath{N2}$. This ratio is defined as
\begin{equation}
    N2\equiv \log \left (\frac{\left[\mathrm{N}\,\textsc{ii} \right]\lambda 6583}{\mathrm{H}\alpha}\right )
.\end{equation}
Then, the metallicity is derived using the following equation:
\begin{equation}
    12 + \log \left (\mathrm{O}/\mathrm{H} \right) = 8.90 + 0.57 \times N2
.\end{equation}

No extinction correction is necessary given the proximity in wavelength of the two lines (\ha\, and \nii$\lambda6583$). To avoid the systematic effect of different abundance calibrations, when comparing the MZR with data from the literature, all the metallicities presented in this work are calculated using the $\ensuremath{N2}$ method. We only calculated the abundance for galaxies marked as star-forming galaxies in the catalogue presented in \cite{glace}, taking out all those classified as AGNs.

We obtained stellar mass estimations for star-forming galaxies by modelling SEDs based on the broad-band ($BVRIJK_{\rm s}$), publicly available optical and near-infrared data described by \cite{moran2005}, and fixing redshift to the values derived from the inverse convolution method. We used a set of low-resolution, composite stellar population templates with delayed exponential star formation histories (SFHs) obtained from the stellar population synthesis models of \cite{bruzual2003}, adopting four stellar metallicity values (from $Z = 0.0004$ to $Z_\odot$; Padova 1994 tracks) and an initial mass function (IMF) according to \cite{chabrier2003}. The age of star formation was constrained to between 10 Myr and 10 Gyr (about the age of the Universe at the mean redshift of the cluster) and the e-folding times in the $ 0.01 < \tau < 30\, {\rm Gyr}$ range. The best fit of the models was obtained using {\tt LePhare} (\citealt{arnouts1999,ilbert2006}), assuming the \cite{calzetti2000} extinction law with intrinsic reddenning $E(B-V)$ varying between 0 and 0.5 mag. The median $\chi^2_{\rm red}$ value obtained from SED fitting is 1.57 and the medians of stellar mass estimations are distributed in the $ 8.4 \le \log\,(M_*\,[M_\odot]) \le 11$ range, with a mean uncertainty of about 0.3 dex. Those estimations are fully consistent with those obtained from exponentially declining SFH and the same prescriptions as those indicated above.

In order to make a meaningful comparison with other MZRs from the literature, we follow the recipe suggested by \cite{mirko2020}, where the data were binned and fitted to the equation adapted from \cite{zahid2014}, which in turn is an evolution of the functional relationship presented in \cite{moustakas2011}:
\begin{equation}
    12 + \log(\mathrm{O}/\mathrm{H}) = \mathrm{Z}_0 + \log\left( 1 - 10^{-\left(\frac{M_*}{M_0}\right)^{\gamma}} \right),
    \label{ecmirko}
\end{equation}

\noindent where according to \cite{mirko2020}, $M_*$ is the stellar mass in solar masses; $Z_0$ is the saturation metallicity, which gives the upper metallicity limit of the MZR; $M_0$ is the turnover mass; and $\gamma$ is the power law that controls the MZR at $M_*<M_0$. This fit is applied to binned data of $\log\, (M_*)$. \cite{mirko2020} used 0.15\,dex stellar mass bins. 
In order to improve the dispersion of the fits, we decided to fix the saturation abundance. According to \cite{marino2013}, the $N2$ ratio begins to saturate at $N2\simeq-0.4$, meaning it no longer presents a linear relationship with oxygen abundance. This is equivalent to 12+$\log$(O/H)$\sim8.67$, and so we assume that this is the asymptotic metallicity of the MZR for galaxies with $M_*>>M_0$. In this way, the parameters to fit are reduced to two: the turnover mass $M_0$, and the exponent of the power law $\gamma$.

To obtain the parameters of the fitting, a set of Monte Carlo simulations were carried out. Each galaxy had a simulated random value of $\log(M_*)$ and of $12+\log(\mathrm{O/H})$ inside the uncertainty of both quantities of the given object. Then, a new MZR was constructed with these simulated values. After that, the MZR was binned in stellar mass. Due to our smaller number of galaxies in clusters at larger redshift, when compared with the \cite{mirko2020} set, our bins need to be larger in order to include a significant number of galaxies. In the end, we employed 0.62\,dex stellar mass bins for our data. Also, this same binning was employed for all the comparison clusters from the literature. This process was repeated 100,000 times, resulting in a set of binned MZRs. Each binned MZR was then fitted using a least-squares minimisation algorithm based
on the Levenberg--Marquardt method. 
The mean and standard deviation were calculated for all the fits. The mean was considered the best fit for the MZR, and the standard deviation was assumed to be the main uncertainty for the MZR. Moreover, the resultant parameters ($M_0$ and $\gamma$) and their uncertainty were calculated as the mean of all the values obtained in every fit and its standard deviation, respectively. 

\begin{figure}[]
    \centering
    \includegraphics[width=\hsize]{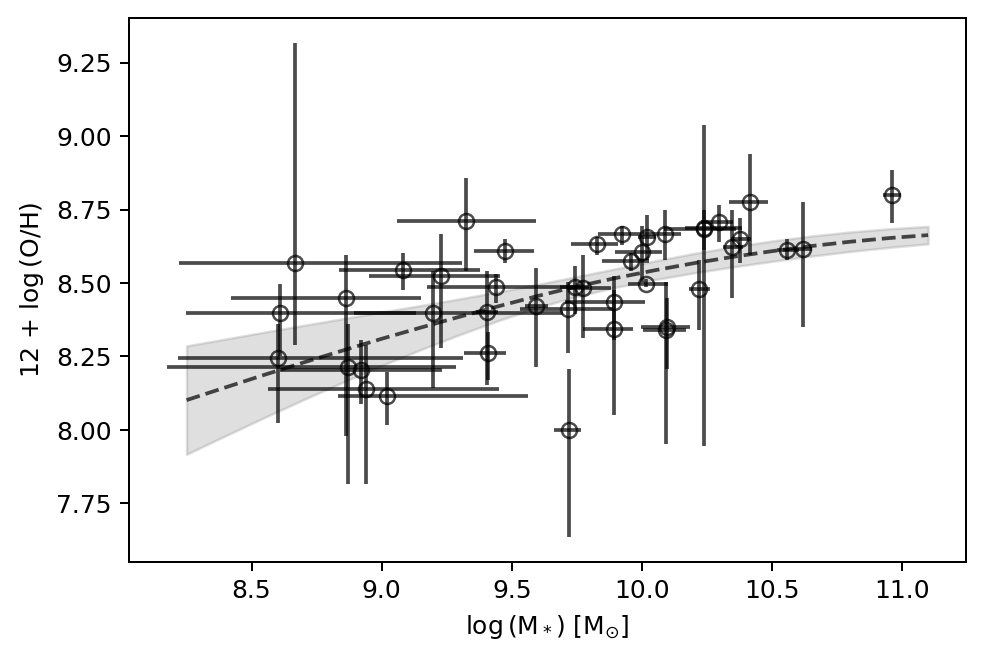}
    \caption{MZR for the SFGs of Cl0024. Our data are the open circles, the black-dashed line represents the mean of the fitted values employing the recipe by \cite{mirko2020}, and the grey area represents the 1$\sigma$ deviation of the fits.}
    \label{mmn2}
\end{figure}

In Fig. \ref{mmn2} we show the MZR for the ELGs detected in the cluster Cl0024. The grey dashed line represents the mean fitted Eq. \ref{ecmirko}. The filled grey area is the 1$\sigma$ uncertainty on the fits. As expected, there is a tight correlation between the stellar mass and the metallicity of the galaxies, with lower-mass galaxies having lower values of oxygen abundance.

In Table \ref{clusters} we summarise the properties of the clusters used in this study. Table \ref{ajustepar} lists the fit parameters of Eq. \ref{ecmirko} for all the clusters, local SDSS galaxies in clusters, and field galaxies at $z\sim 0.4$, which we use for comparison in the following sections. 

\begin{table*}[]
    \centering
    \caption{Main characteristics of the galaxy clusters employed in this study.}
    {\renewcommand{\arraystretch}{1.3}%
    \begin{tabular}{lcccc}
    \hline
    \hline
     Cluster name   & Redshift  & $M_{200}$ [$M_{\odot}$] & $R_{200}$ [Mpc] & $<\!Z/Z_{\odot}\!>$\\
     \hline
     Hercules Supercluster & 0.033 & $2.1 \times 10^{15}$(1) & N/A & N/A\\
     RX\, J2248--443 & 0.348 & $2.81 \times 10^{15}$ (2) & 2.6 (3) & 0.26 (4)\\
     Cl\,0024 (this work) & 0.395 & $5.85 \times 10^{14}$ (5) & 1.73 (5) & 0.22 (6)\\
     MACS\,J0416.1--2403 & 0.397 & $1 \times 10^{15}$ (7) & 1.8 (7)& 0.24 (8)\\
     Cl\,0939+4713 & 0.41 & $1.7 \times 10^{15}$ (3) & 2.13 (3) & 0.2 (9)\\
     XMMXCS\,J2215.9--1738 & 1.5 & $6.3 \times 10^{14}$ (10)& 1.23 (10) & N/A\\
     XMM--LSS\, J02182--05102 & 1.62 & $7.7 \times 10^{13}$
     (11) & 0.49 (11) & N/A\\
     PKS\,1138--262 & 2.16 & $1.71 \times 10^{14}$ (12) & 0.53 (12) & N/A\\
     \hline
    \end{tabular}}
    \tablefoot{First column is the cluster name, second column is the redshift of the cluster, third column is the $M_{200}$ in solar masses, fourth column is the $R_{200}$ in Mpc, and fith column is the mean value of the iron abundance in the intracluster gas in units of solar metallicity.}
    \tablebib{ (1)~\citet{monteiro2022}; (2) \citet{kesebonye2023}; (3) \citet{koyama2011}; (4) \citet{defilippis2003}; (5) \citet{glace};  (6) \citet{zhang2005}; (7) \citet{bonamigo2018}; (8) \citet{bonamigo2017}; (9) \citet{rahaman2021}; (10) \citet{Maier2019}; (11) \citet{pierre2012};  (12) \citet{shimakawa2014}.}
    \label{clusters}
\end{table*}

\begin{table}[]
    \centering
    \caption{Parameters fitted in Eq. \ref{ecmirko} for different clusters and field galaxies.}
    {\renewcommand{\arraystretch}{1.3}%
    \begin{tabular}{lcc}
        \hline
        \hline
        Cluster  & {$\log(M_0/M_{\odot})$} & {$\gamma$} \\
        \hline
        SDSS local clusters & 9.50$\pm$0.17 & 0.32$\pm$0.07\\
        Hercules Supercluster & 9.72$\pm$0.22 & 0.49$\pm$0.05\\
        RX J2248–443 &  9.32$\pm$0.45 & 0.41$\pm$0.04 \\
        Cl0024 (This work) & 10.68$\pm$0.34 & 0.36$\pm$0.09\\
        MACS J0416.1--2403 & 10.46$\pm$0.05 & 0.42$\pm$0.08\\
        Field galaxies at $z\sim0.4$ & 11.66$\pm$0.51 & 0.21$\pm$0.05\\
        Cl 0939+4713 &  9.42$\pm$1.0 & 0.12$\pm$0.08\\  
        XMMXCS J2215.9--1738 & 10.59$\pm$0.06 & 0.63$\pm$0.17\\
        XMM--LSS J02182--05102 & 11.08$\pm$0.52 & 0.32$\pm$0.07\\
        PKS 1138–262  & 10.48$\pm$0.26 & 0.30$\pm$0.08\\   
        \hline
    \end{tabular}} 
    \tablefoot{First column is the cluster name; second column is the logarithm of the turnover mass expressed in solar masses; column 3 is the exponent of the power law $\gamma$.}
    \label{ajustepar}
\end{table}

\subsection{The effect of the environment}
\begin{figure*}
  \subfloat{\includegraphics[width=0.5\hsize]{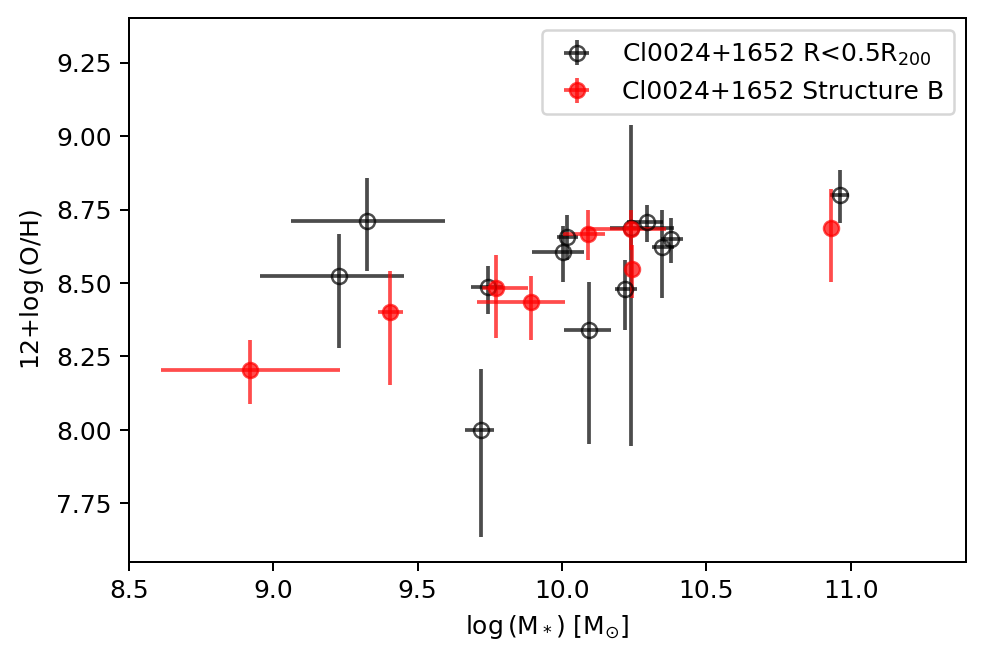}}
 \subfloat{\includegraphics[width=0.5\hsize]{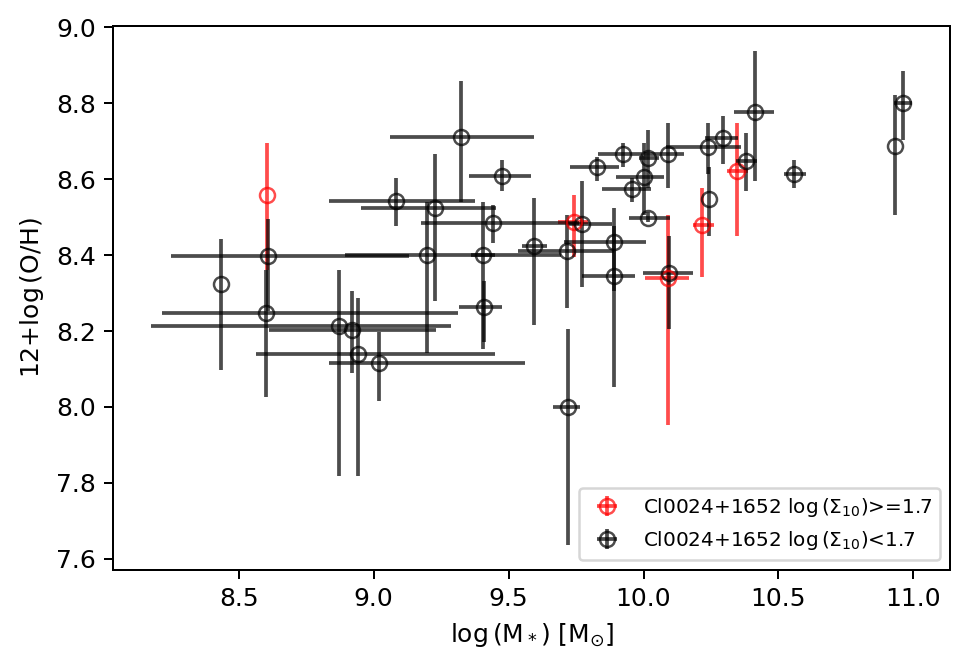}}
    \caption{MZR for Cl0024. Left panel: Open black circles and red filled circles are the galaxies with $R<0.5\, R_{200}$ and galaxies in Structure B, respectively. Right panel: Open red circles and open black circles are the galaxies from Cl0024 within medium and high-density zones ($\log(\Sigma_{10})\geqq 1.7$) and low-density zones ($\log(\Sigma_{10})<1.7$), respectively.}
    \label{mzr_env}
\end{figure*}
Figure \ref{mzr_env}  shows the MZR for the galaxies of Cl0024.  In the left panel, we only show the galaxies of the main structure inside $0.5\,R_{200}$ (black open circles) and the galaxies of the infall group (red circles). In the right panel, we separate Cl0024 galaxies into medium and high-density zones, $\log(\Sigma_{10})\geqq1.7$, following the criteria suggested by \cite{sobral2016}; these are represented by open red circles, and galaxies in low-density zones are represented by open black circles. In order to eliminate some of the noise in the figure, we only consider the galaxies with an uncertainty of 1\% in the abundance determination.\\

\cite{Maier2019} found that the galaxies with $0.5\,R_{200}$ have enhanced metallicities when compared to the infall galaxies of the same cluster. However, we find no such differences between the infall group and the galaxies inside $0.5\,R_{200}$. An explanation for this difference could be that XMMXCS J2215.9--1738 infall galaxies come from the field, and are therefore poorer in metals when compared with the galaxies from the core. On the other hand, the infall group in Cl0024 is, in reality, a subgroup of galaxies with a certain density above that of the field galaxies.\\ 

The right panel of Fig. \ref{mzr_env}  also shows that there is no difference between galaxies in high-density zones and those in low-density zones. All these results seem to agree with \cite{sobral2016}, who found no significant dependence of the MZR on the environment inside the same cluster.

\begin{figure}[]
    \centering
    \includegraphics[width=\hsize]{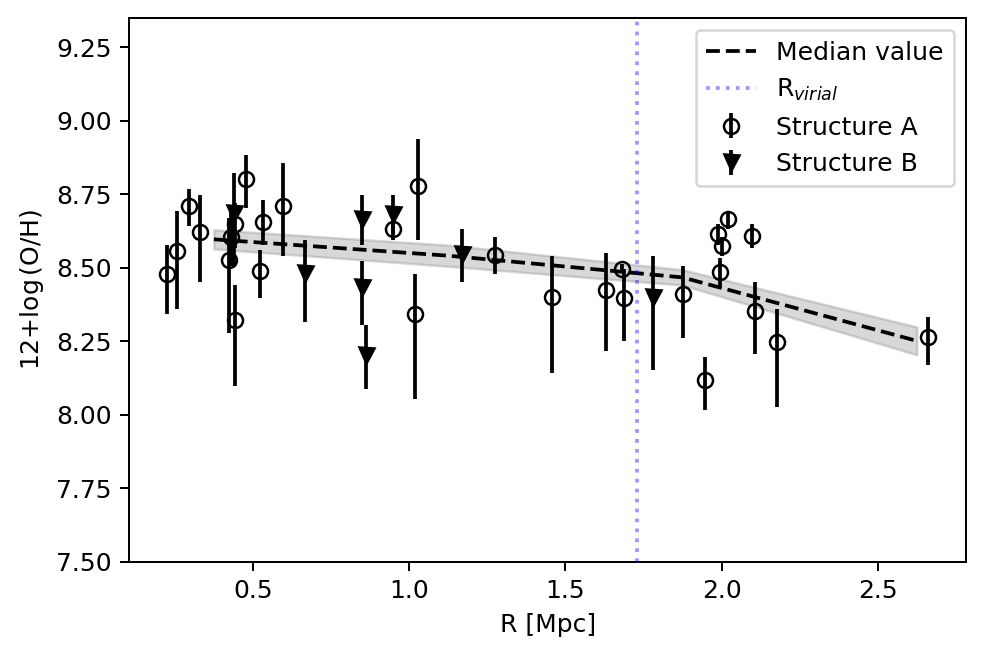}
    \caption{Metallicity as a function of clustercentric radius for Cl0024. The open black circles are the galaxies from Structure A. The downward black-filled triangles are the galaxies from Structure B. The grey dashed line is the median of the values for Structure A. The blue dotted line indicates the position of the virial radius for Cl0024 from \cite{glace}.}
    \label{mzrradio}
\end{figure}
 Figure \ref{mzrradio} shows the abundance as a function of the clustercentric radius for galaxies on Structures A and B. The median value is represented by a grey dashed line and only for galaxies in Structure A. This median value and its uncertainty were calculated in the same way as the binned median MZR was computed, as described in section \ref{sec:mzr}. The filled grey zone represents the uncertainty on the median value of $1\sigma$. The blue dotted line represents the position of the $R_{\mathrm {virial}}$ from \cite{glace}.
There is a slight dependence of the metallicity on clustercentric radius, with inner galaxies presenting, at least on average, larger abundances. We find no clear differences between galaxies in Structures A and B. However, \cite{maier2015} found that the metallicities of galaxies inside the virialized part of the cluster they studied presented enhanced metallicities by about 0.1\,dex when compared to infalling galaxies. Here, the difference in median abundance between galaxies well inside the virialized part and galaxies in the outside zones is about $\sim0.25$\,dex, which is slightly larger than the difference reported by \cite{maier2015}. This is a result similar to the one presented in \cite{ellison2009}, where it is reported that the galaxies are slightly more metal-rich if they are located in overdensity zones. \cite{maritza2022} also find a dependence of the metallicity on projected clustercentric radius for galaxies in the Fornax cluster. 
On the other hand, \cite{mouhcine2007} and \cite{cooper2008} found weak or no connection between abundance and environment in galaxies. However, \cite{cooper2008} expanded the study, and after removing the mean colour--luminosity--environment relation, found a residual relationship between environment and metallicity.

\subsection{Effect of the mass of the cluster}\label{mass_cluster}
\begin{figure}[]
   \centering
   \includegraphics[width=\hsize]{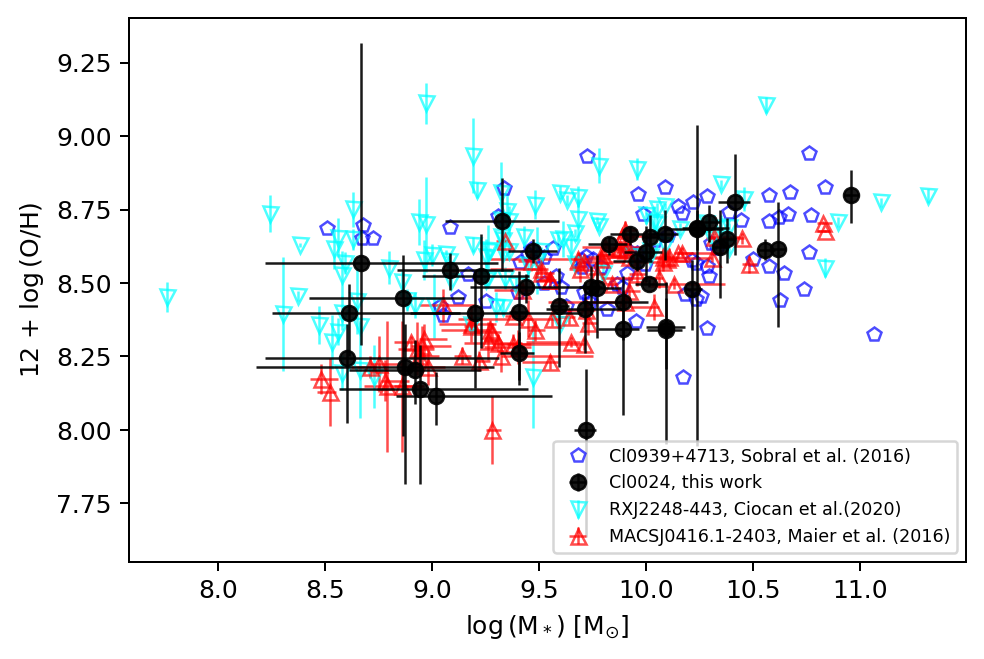}
      \caption{Mass--metallicity relationship for ELGs in clusters at similar redshift. Black circles represent our data. The cyan downward-pointing triangles are the galaxies from \cite{ciocan2020} for the cluster RX J2248-4431. The blue pentagons represent the galaxies from \cite{sobral2016} for the cluster Cl 0939+4713. The red upward-pointing triangles are the galaxies from \cite{maier2016} for the cluster MACSJ0416.1--2403.}
         \label{masmet1}
\end{figure}

In Fig. \ref{masmet1} we make a comparison with data from cluster RX J2248-4431 (thereafter RXJ2248) from \cite{ciocan2020}, cluster MACS J0415.1--2403 (thereafter MACS0415) from \cite{maier2016}, and cluster Cl0939+4713 (thereafter Cl0939) from \cite{sobral2016}. All clusters are at a similar redshift.
\begin{figure}[]
    \centering
    \includegraphics[width=\hsize]{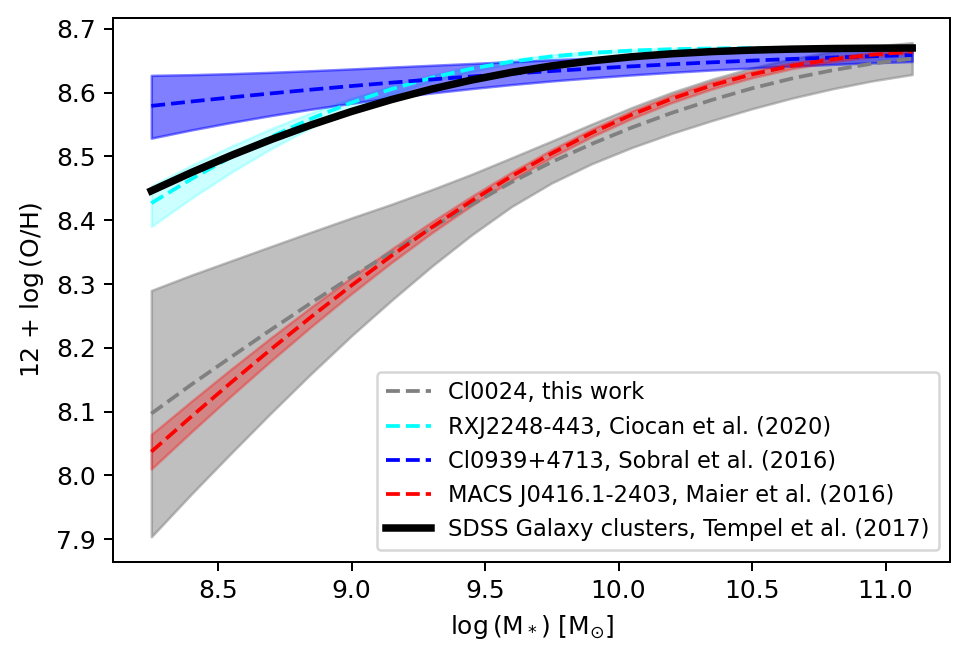}
    \caption{Comparison of MZR between Cl0024 and clusters at the same redshift, as well as galaxies from clusters at low redshift from \cite{tempel}. The colour coding is reported in the legend. All data were fitted following the recipe presented in \cite{mirko2020}.}
    \label{binmzr}
\end{figure}

 Figure \ref{binmzr} shows the fitted functions described in Eq. \ref{ecmirko} with one standard deviation (filled regions) of the MZR for the four clusters and the SDSS galaxy clusters (black thick line) from \cite{tempel} at local redshift. It should be noted that all four clusters have been chosen in such a way that they are in a similar non-relaxed status (i.e. \citealt{glace} and \citealt{czoske2002} for Cl0024; \citealt{schindler1998} for Cl0939; \citealt{mann2012} for MACS0416; and \citealt{rahaman2021} and \citealt{kesebonye2023} for RXJ2248). 
We can see that the fitted value of the abundance for RXJ2248 is systematically larger than the abundance for our data of Cl0024. On the other hand, the data from MACS0416 follow, inside uncertainties, the same path as Cl0024. The data from Cl0939 seem to follow a shallower path when compared with RXJ2248. However, the fact that the data from \cite{sobral2016} do not include uncertainties should be taken into account. In the end, we decided to add a $5\%$ error to each measured flux. Therefore, it is possible that, in this case, we are underestimating the total uncertainty of the fit. Nevertheless, the value of the abundance for Cl0939 is larger than Cl0024 for all ranges in stellar mass. Moreover, most of the time, the abundance fitted for RX J2248 and Cl0939 is larger than or equal to that in local clusters, and the abundances for Cl0024 and MACS0416 are below them. If we take into account the fitted parameters presented in Table \ref{ajustepar}, we can see that the main difference between the clusters lies in the turnover mass, $\log(M_0/M_{\odot})$, with Cl0024 and MACS0416 having a large value (10.68 and 10.46 respectively) when compared to Cl0939 and RXJ2248 clusters. This indicates that $\gamma$, the low-mass end slope, is dominant for almost all stellar masses sampled in Cl0024 and MACS0416. The shallow behaviour of Cl0939, represented by the low value of $\gamma=0.12\pm0.008$, may be due to the lack of low-stellar-mass galaxies when compared with the other two clusters, which generates a poor constraint when fitting. On the other hand, the $\gamma$ obtained for Cl0024, RXJ2248, MACS0416, and local clusters are similar within the uncertainties.

From Table \ref{clusters}, it can be seen that the $M_{200}$ masses for RXJ2248 and Cl0939 are $2.81 \times 10^{15}M_{\odot}$ and $1.7 \times 10^{15}M_{\odot}$, respectively. Both are an order of magnitude higher than the reported mass for Cl0024. On the other hand, the $M_{200}$ for MACS0416 is somewhat lower, less than two times larger than the mass for Cl0024. 

According to \cite{calabro2017}, the MZR is a consequence of the conversion of gas into stars inside galaxies, a process regulated by gas exchanges with the environment. There should therefore be a correlation between the total abundance of the gas in the galaxies and the metallicity of the intracluster gas. However, in Table \ref{clusters} we can see that the average values of $Z/Z_{\odot}$ for Cl0024, Cl0939, MACS0416, and RXJ2248 are similar. 

Nevertheless, we have to take into account that the metallicity is not constant within each cluster, and here we are comparing average values. According to \cite{defilippis2003}, Cl0939 shows a high-metallicity region (they called it M1, with $Z/Z_{\odot}\simeq0.33$) with a peak in galaxy number density. These authors suggest that the galaxies there may have enriched the gas in the zone compared with others with a lower galaxy number density. For RXJ2248, \cite{rahaman2021} gives $Z/Z_{\odot}\simeq0.36$ for the centremost zone of the cluster. On the other hand, \cite{zhang2005} give a value of $Z/Z_{\odot}\simeq0.25$  for the most metallic
zone in Cl0024. If we take this into account, it is clear that Cl0939 and RXJ2248 show zones where the intracluster gas has larger abundances than the most metallic zone of Cl0024. 

\begin{figure}
    \centering
    \includegraphics[width=\hsize]{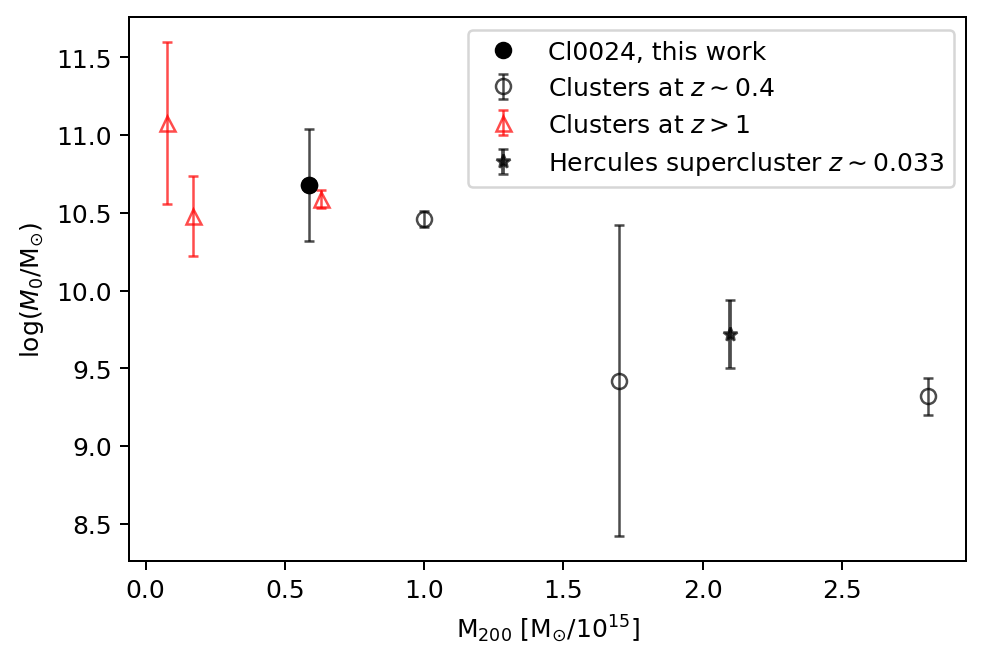}
    \caption{Fitted turnover parameter from Eq. \ref{ecmirko} for all the clusters in the study as a function of M$_{200}$. The black-filled and open circles are, from left to right, the clusters at $z\sim 0.4$, Cl0024, MACS0416, Cl0939, and RXJ2248, respectively. The open red triangles are, from left to right, the clusters at $z>1$,  XMM--LSS J02182--05102, PKS\,1138--262, and XMMXCS J2215.9--1738. The black star is the Hercules Supercluster at $z\sim0.033$.}
    \label{turn}
\end{figure}

Figure \ref{turn} shows the logarithm of $M_0$ as a function of M$_{200}$ for all the clusters presented in Table \ref{clusters}. The filled black circle is the cluster Cl0024, and the open black circles are the clusters at about $z\sim0.4$. The open red triangles are the clusters at larger redshift. The filled black star represents the Hercules Supercluster at $z\sim0.033$. The figure indicates that the drive for the change in the turnover mass, and therefore in the larger abundance for several clusters, lies in the M$_{200}$ of the cluster: the more massive the cluster, the higher the density of galaxies in certain zones, even if the mean value of the metallicity of the intracluster gas is the same. 

\subsection{Effect of the redshift}\label{redshift}

\begin{figure}
    \centering
    \includegraphics[width=\hsize]{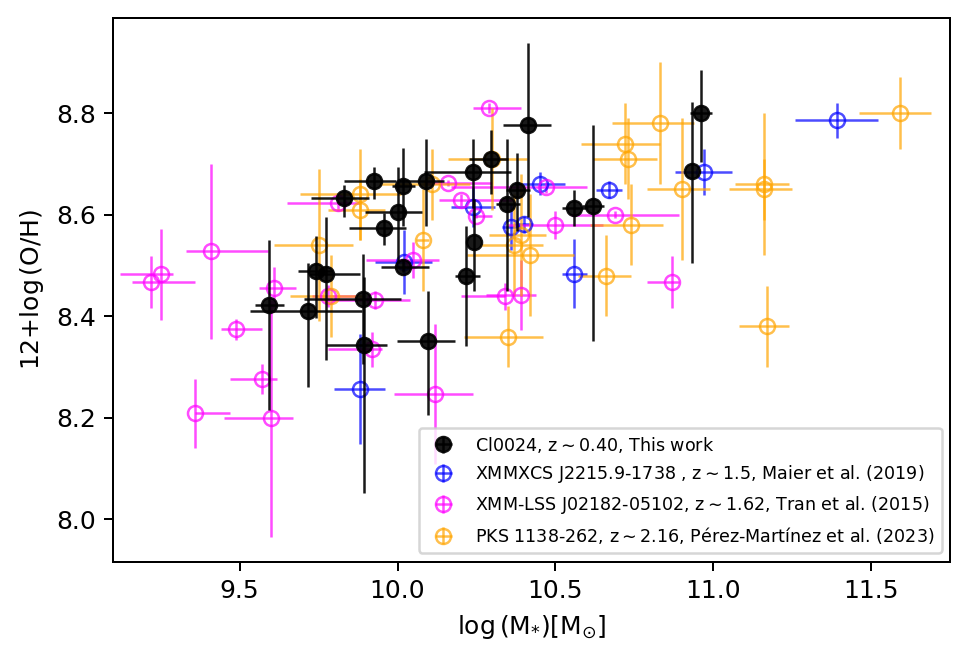}
    \caption{MZR for ELGs in clusters at different redshifts. We show the galaxies from Cl0024 with errors in the determination of the metallicity below 0.3\,dex and stellar masses over $\log(\mathrm{M}/M_{\odot})=9.5$ with filled black circles, the data from \cite{Maier2019} for the cluster XMMXCS J2215.9--1738 at $z\sim1.5$ with blue circles, the data from \cite{tran2015} for XMM--LSS J02182--05102 at $z\sim1.62$ with magenta circles, and the data from \cite{perez2023} for the cluster PKS 1138--262 at $z\sim2.16$ with orange circles.}
    \label{mzrredshiftpoint}
\end{figure}
\begin{figure}
    \centering
    \includegraphics[width=\hsize]{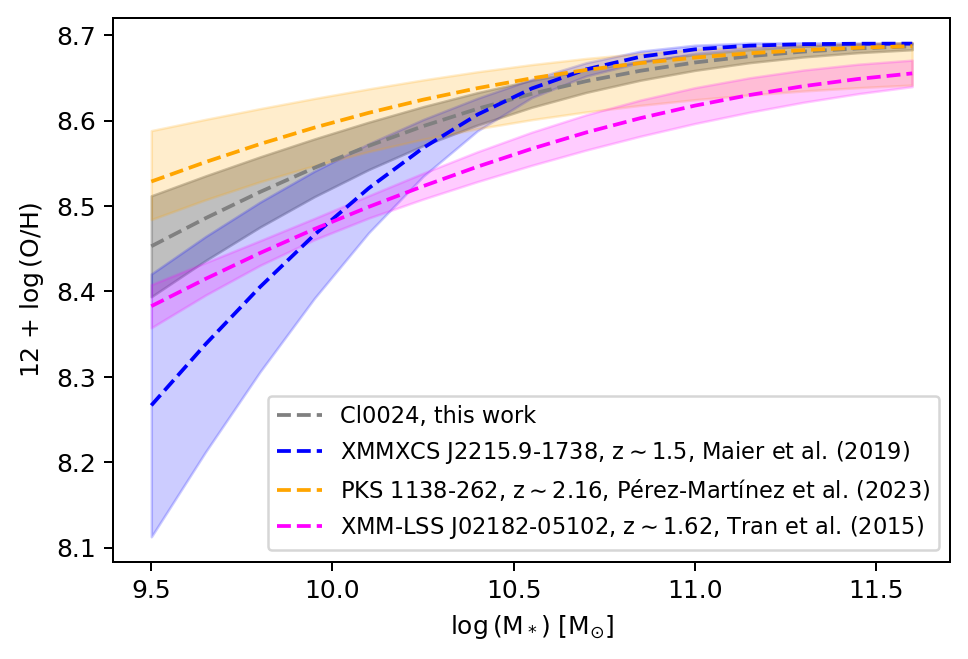}
    \caption{MZR for ELGs in clusters at different redshifts. We show the fitted values from Eq. \ref{ecmirko} and the 1$\sigma$ deviation for the three clusters (filled regions) employing the same colours as in Fig. \ref{mzrredshiftpoint}.}
    \label{mzrredshiftfit}
\end{figure}

Figure \ref{mzrredshiftpoint} shows the MZR for clusters at different redshifts: cluster XMMXCS J2215.9-1738, hereafter XCS2215, at $z\sim$1.5 from \cite{Maier2019}, cluster XMM--LSS J02182--05102, hereafter LSS2182, at $z\sim1.62$ form \cite{tran2015}, and the cluster PKS 1138-262, hereafter PKS1138, at $z\sim$2.16 from \cite{perez2023}.  XCS2215 and PKS1138 have similar values for $M_{200}$. On the other hand, LSS2182 has a somewhat lower value of $M_{200}$, but with large uncertainty, $(7.7\pm3.8)\times10^{13}M_{\odot}$ (\citealt{pierre2012}), and so we can consider that, in this case, we are mainly comparing the effect of the redshift between the three clusters. In order to carry out a meaningful comparison with the high-redshift clusters, which are limited to high-stellar-mass galaxies, we restricted the galaxies of Cl0024 to stellar masses larger than $\log(M_*/M_{\odot})>9.5$.

Figure \ref{mzrredshiftfit} shows the fit drawn using Eq. \ref{ecmirko}. All the clusters are approximately superimposed. The fitted values for PKS1138 and Cl0024 are the same within uncertainties. From Table \ref{ajustepar}, we can see that the values of the turnover mass are also compatible within the uncertainties, as shown in Fig. \ref{turn}. The only difference between the three clusters is on the $\gamma$ parameter, which in this case is somewhat higher for XCS2215 when compared with the other clusters.

For a fixed stellar mass, \cite{ly2016} show that the abundance of field galaxies evolves with redshift as $\log(\mathrm{O}/\mathrm{H})\propto(1+z)^{-2.32}$. Moreover, \cite{maiolino2008} found an evolution on the MZR with redshift, and found it to be faster at $2.2<z<3.5$ than at lower redshifts ($z<2.2$). 
Also \cite{huang2019} also found that the more evolved galaxies have higher metallicities at fixed stellar masses when compared with less evolved galaxies.  Also, \cite{troncoso} found that at $z\sim3.4$, galaxies are more metal poor when compared with lower-redshift galaxies. We do not find a similar relationship with galaxy clusters, although we do not have data over $z>2$. This may imply that the effect of the environment, such as the total mass of the cluster, may have a larger influence than the canonical evolution of the cluster.

\subsection{Comparison with field galaxies at the same redshift}

\begin{figure}
    \centering
    \includegraphics[width=\hsize]{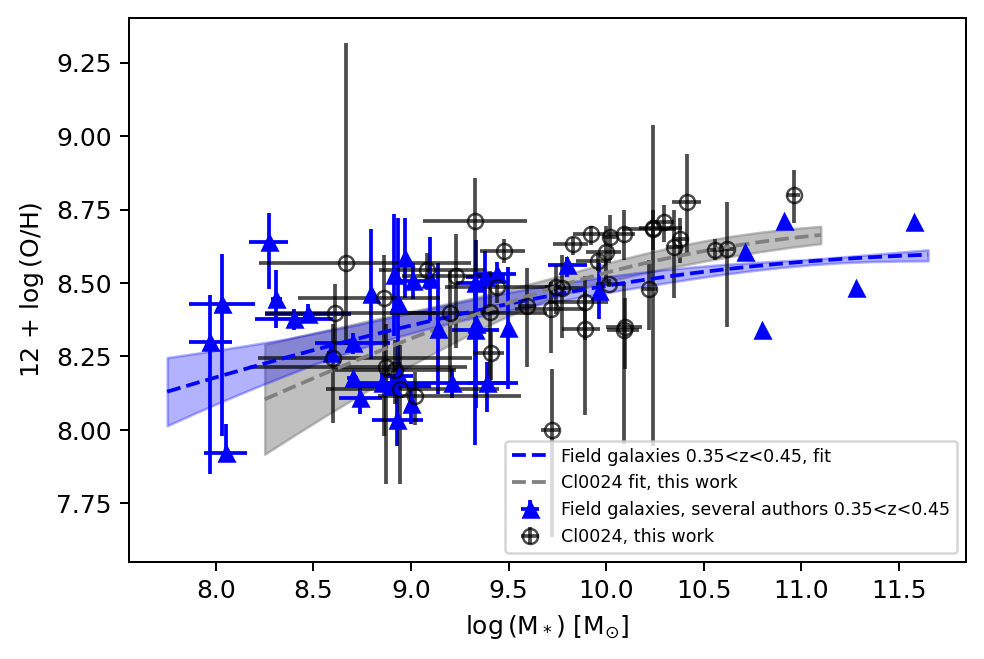}
    \caption{MZR for Cl0024 and field galaxies at similar redshift. Cl0024 galaxies are represented as open grey circles, and field galaxies are represented as filled blue triangles. The field galaxies come from \cite{jakub2020}, \cite{amorin2015}, and SDSS galaxies at $0.35<z<0.45$. The fitted value for the MZR is represented by the grey dashed line and the blue dashed line for Cl0024 and field galaxies, respectively. The filled areas represent a 1$\sigma$ deviation of the fit.}
    \label{mzr_samered}
\end{figure}

Figure\,\ref{mzr_samered} shows the MZR for our galaxies in Cl0024 and field galaxies at the same redshift from several authors: \cite{amorin2015}, \cite{jakub2020}, and SDSS field galaxies at $0.35<z<0.45$ from the \cite{tempel} catalogue, as well as SFGs with  $0.35 < z < 0.45$ from VIMOS VLT Deep Survey Database (see \citealt{vimos2005} for the description of the survey). From Table \ref{ajustepar}, we can see that the turnover mass for the field galaxies is the largest one: $\log(M_*/M_{\odot})=11.66\pm0.51$ when compared with $\log(M_*/M_{\odot})=10.68\pm0.34$ for Cl0024 or $\log(M_*/M_{\odot})=9.50\pm017$ for SDSS local clusters. This means that the power-law part of Eq. \ref{ecmirko} is dominant for a longer range.

In this case, for lower stellar masses, the abundance of Cl0024 is compatible with the field galaxies at the same redshift, although the abundance is slightly lower for Cl0024. However, at masses over $\log(M_*/M_{\odot})>9.75$, the metallicity of the cluster is slightly larger than that of the field galaxies. This seems to agree with the results from \cite{kacprzak2013}, where the authors found no difference larger than 0.02\,dex in the MZR of field galaxies at $z\sim2$ when compared with cluster galaxies at the same redshift.

\section{Discussion and conclusions}

\begin{figure}
    \centering
    \includegraphics[width=\hsize]{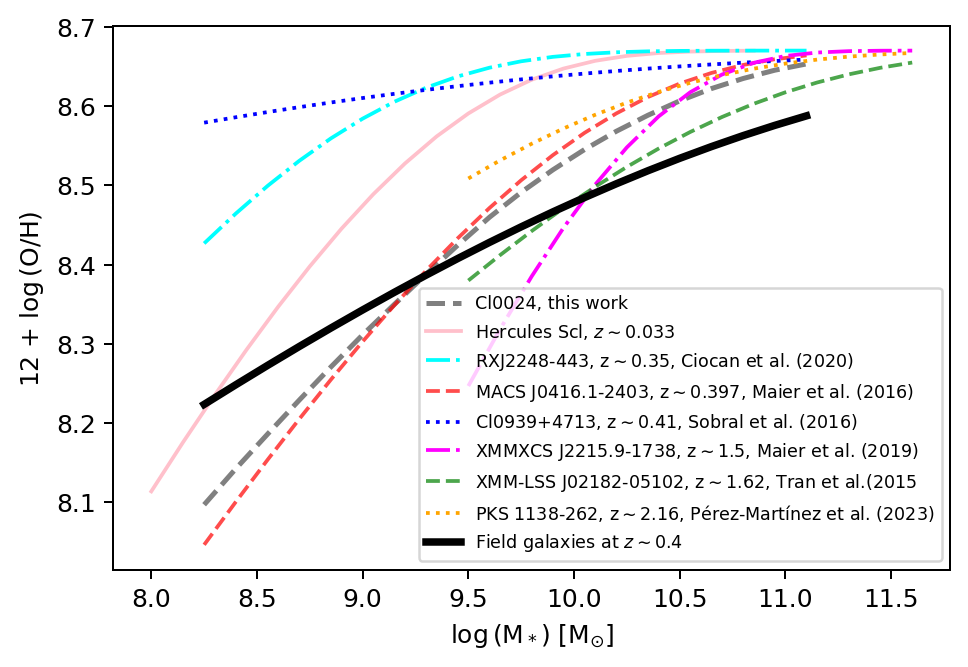}
    \caption{MZR fits for galaxy clusters and field galaxies at $z\sim0.4$. The meaning of the colour code and shape of the lines is explained in the inset key.} 
    \label{allcon}
\end{figure}

\begin{figure}
    \centering
    \includegraphics[width=\hsize]{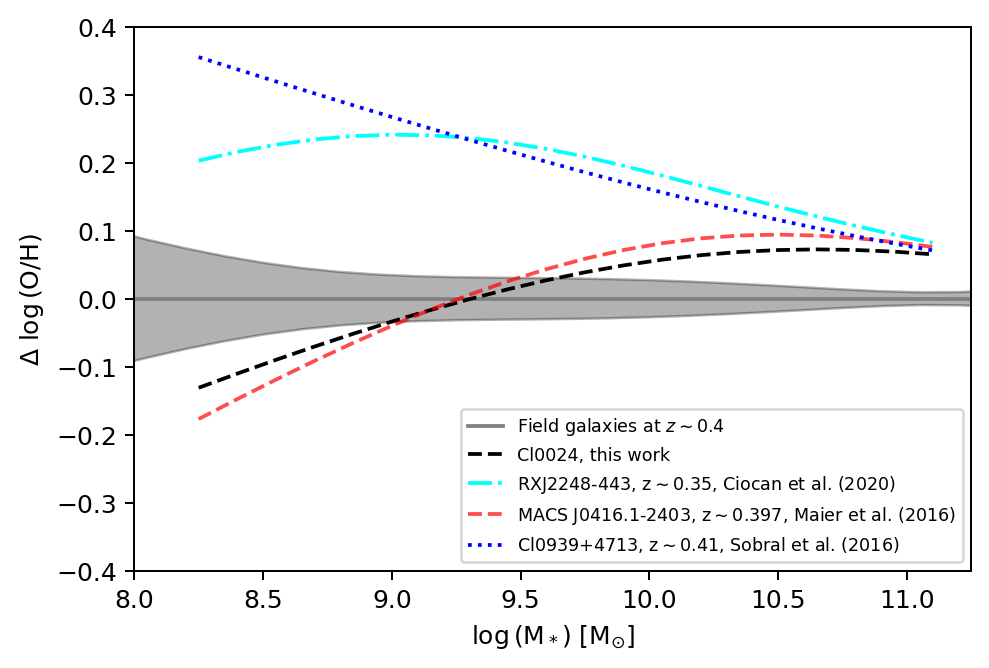}
    \caption{Offsets from the fitted value of MZR of field galaxies at $z\sim0.4$ for clusters at similar redshift. The filled area represents a 1$\sigma$ deviation of the fit for field galaxies. The colour coding and shape of the lines are explained in the inset key.}
    \label{variamet}
\end{figure}

 Figure \ref{allcon} shows all the fits calculated for Eq. \ref{ecmirko} for all the clusters presented in this work and field galaxies at $z\sim0.4$ from several sources. We also include data from the Hercules Supercluster, a massive cluster (M$_{200}=2.1\times10^{15}M_{\odot}$ according to \citealt{monteiro2022}) at an {almost local} redshift of $z\sim 0.033$. These data were obtained from \cite{vasiliki2011} for Abell 2151 (the Hercules cluster proper) and from \cite{vasilikitesis} for the rest of the subclusters that make up the Hercules Supercluster.

Figure \ref{variamet} shows the offsets from the fit of the MZR of field galaxies at $0.35<z<0.45$. It is clear that Cl0024 and MACS0416 clusters both have lower metallicity at lower stellar masses when compared with field galaxies; they also present lower metallicities at all stellar masses when compared with the two more massive clusters Cl0939 and RXJ2248. It should be noted that the two massive clusters RXJ2248 and Cl0939 present a larger value of the turnover parameter $M_0$ when compared with the rest of the clusters. The same also happens at local values of the redshift for the Hercules Supercluster (see Fig. \ref{turn}).

The difference presented in Fig. \ref{variamet} can be explained if we consider that more massive clusters should present larger zones with high density, where the galaxy--galaxy encounters as well as galaxy--cluster interactions (due to a large virial radius) may be more effective than in smaller clusters. For this reason, low-mass galaxies immersed in massive clusters tend to be more metallic than those in smaller clusters. Therefore, in all these cases, the value of the $M_0$ parameter may be a proxy for the mass of the cluster, and therefore an indicator of the quenching of the star formation in clusters. However, a proper study of the SFR is being carried out (De Daniloff
et al. in prep.) to confirm whether or not we are seeing a genuine quenching of the star formation.

Regarding the dynamical state of the cluster, gas metallicity estimations appear not to be affected by the membership of the galaxy to any of the two identified structures confirmed by \cite{glace}. So far, we can simply conclude that such structural duality only appears to affect density indicators, in the sense that the median projected surface density ($\Sigma_{10}$) of Structure A is $\sim$0.7\, dex denser than that of Structure B (see Section \ref{sec:environment}).

On the other hand, as seen in Figs. \ref{mzr_samered} and \ref{variamet}, the less massive clusters (Cl0024 and MACS0416) present values for MZR closer to those of field galaxies at the same redshift, with differences in abundance at lower masses of about -0.15\,dex and 0.1\,dex when compared to the more massive galaxies. This seems to indicate that the conditions in these clusters are much closer to those of field galaxies than to those of more massive clusters.

\subsection{Conclusions}
The main conclusions we can derive from this work are as follows:
\begin{itemize}
    \item We revisited the work of \cite{glace} and applied the inverse convolution method to the galaxy cluster Cl0024. We obtain a total of 84 galaxies (classified in tiers 1 to 3) from the original 174, where the method yields better results in the determination of the fluxes and redshifts, and we are able to reduce the uncertainties from over 20\% in the old catalogue to less than 10\% for the 40\% of ELGs, and even below 5\% for the 20\% of the ELGs. 
    \item From the redshift distribution, we find the two clearly differentiated structures previously mentioned in \cite{czoske2002} and \cite{moran2007}, as well as the third interacting structure at about $z\sim0.42$ suggested in \cite{glace}.
    \item There is a relationship between the clustercentric radius and the logarithm of the density, with galaxies with larger radii being located in zones with lower density when compared with galaxies in the central zones. This effect happens for the two main structures. This seems to imply that both structures may have their nucleus in the line of sight, and therefore may be superimposed.
    \item We obtain the MZR for the ELGs of Cl0024 after removing AGNs, indicated as such in \cite{glace}.
    \item We fitted Cl0024 MZR with the function proposed by \cite{mirko2020}, as well as several other clusters and field galaxies.
    \item The MZR for Cl0024 presents a tight correlation between the metallicity and the stellar mass of the galaxies, with galaxies with higher masses having higher values of oxygen abundance.
    \item We separated the galaxies of Structure A and Structure B in the MZR of Cl0024, as well as the galaxies in high-density zones and low-density zones, and find no differences between them. These results agree with \cite{sobral2016}, where we find no significant dependence of the MZR on the environment inside the same cluster.
    \item We find a slight gradient of the metallicity with the clustercentric radius for galaxies of Structure A. Galaxies tend to be more metallic in the inner part of the cluster when compared with galaxies on the outskirts.
    \item We find a difference in MZRs for galaxy clusters at the same redshift. Cl0939 and RXJ2248 data from \cite{sobral2016} and \cite{ciocan2020}, respectively, present larger values of abundance up to stellar masses over $\log(M_*/M_{\odot})>10.5$. This difference seems to depend on the M$_{200}$ of each cluster,
    in the sense that clusters with larger masses have lower values of the turnover mass, $M_0$. This is the case even if the intracluster gas abundance is similar for the clusters. However, the clusters Cl0939 and RXJ2248 present high-density zones with larger values of intracluster gas abundance when compared with Cl0024 and MACS0416. This may imply that the more massive clusters present more galaxy--galaxy as well as cluster--galaxy encounters, which may cause a quenching of the star-forming processes.
    \item No appreciable difference was found in the MZR of clusters of similar or lower M$_{200}$ but different redshift than Cl0024. This seems to suggest that the way in which clusters evolve is heavily influenced by the total mass of the cluster.
    \item When comparing Cl0024 with field galaxies at the same redshift, it we find that the  MZR fit  of  field galaxies presents a larger turnover mass, although there is little difference in the abundance of lower stellar masses. 
\end{itemize}
\begin{acknowledgements}

B.C., M.S.P. and A.B. acknowledge the support of the Spanish Ministry of Science, Innovation and Universities through the project PID-2021-122544NB-C43. J.C. and M.G.-O. acknowledges the support of the Spanish Ministry of Science, Innovation and Universities through the project PID-2021-122544NB-C41. M.A.L.L. acknowledges support from the Spanish grant PID-2021-123417OB-I00, and the Ramón y Cajal program funded by the Spanish Government (RYC2020-029354-I).
MC acknowledges the support of PID2019-107408GB-C41 and PID2022-136598NB-C33 grants funded by MCIN/AEI/10.13039/501100011033 and by “ERDF A way of making Europe”.
B.C. wishes to thank Carlota Leal \'Alvarez by her support during the development of this paper.
Based on observations made with the Gran Telescopio Canarias (GTC), installed at the Spanish Observatorio del Roque de los Muchachos of the Instituto de Astrofísica de Canarias, on the island of La Palma.
This research uses data from the VI-MOSS VLT Deep Survey, obtained from the VVDS database operated by Cesam Laboratoire d'Astrophysique de Marseille, France.
This research has made use of NASA’s Astrophysics Data System.
This research has made use of the NASA/IPAC Extragalactic Database (NED), which is funded by the National Aeronautics and Space Administration and operated by the California Institute of Technology.

\end{acknowledgements}

\bibliographystyle{aa} 
\bibliography{biblio} 

\begin{appendix}
\clearpage
\onecolumn
\section{List of galaxies from tiers 1 to 3}

{\fontsize{9}{11}\selectfont
\begin{longtable}{ccccccc}
    \caption{Catalogue of deconvolved galaxies with quality tiers 1 to 3. The first column is the identification number from \cite{glace}; column 2 is the new derived redshift; column 3 and 4 are the integrated \ha\ flux and \nii$\lambda6583,$ respectively, in erg\,cm$^{-2}$\,s$^{-1} \times 10^{-17}$; column 5 is the derived abundance employing the N2 index; column 6 is the logarithm of the stellar mass in solar masses, and column 7 is the flag for AGN from \cite{glace}. The uncertainties are taken as one standard deviation of the value of each parameter.}\\
    \hline
    \hline
ID (\#) & $z$  & $F$(\ha) & $F$(\nii$_{\,\lambda6583}$) & $12+\log(\mathrm{O}/\mathrm{H})$ & $\log(M_{*})$ & isAGN \\
 & &  [$10^{-17}$ erg\,cm$^{-2}$\,s$^{-1}$] & [$10^{-17}$ erg\,cm$^{-2}$\,s$^{-1}$] & & [$M_{\odot}$]  & \\
\hline
\endhead
\hline
\endfoot
\endlastfoot
73 & 0.3966 $\pm$ 0.0004 & 9.85 $^{+1.69}_{-1.46}$ & 4.44 $^{+0.98}_{-1.19}$ & N.A. & 9.30 $^{+0.28}_{-0.15}$ & True\\
87 & 0.3801 $\pm$ 0.0003 & 12.34 $^{+2.22}_{-2.13}$ & 1.33 $^{+1.18}_{-1.33}$ & N.A. & N.A. & True\\
105 & 0.3829 $\pm$ 0.0002 & 31.60 $^{+4.29}_{-3.91}$ & 4.21 $^{+2.46}_{-2.16}$ & 8.40 $^{+0.14}_{-0.25}$ & 9.40 $^{+0.04}_{-0.04}$ & False\\
106 & 0.3810 $\pm$ 0.0002 & 25.74 $^{+3.22}_{-3.12}$ & 9.98 $^{+2.39}_{-2.14}$ & 8.67 $^{+0.08}_{-0.09}$ & 10.09 $^{+0.06}_{-0.09}$ & False\\
138 & 0.3814 $\pm$ 0.0002 & 42.46 $^{+4.89}_{-4.58}$ & 7.87 $^{+2.78}_{-3.14}$ & 8.48 $^{+0.11}_{-0.17}$ & 9.77 $^{+0.11}_{-0.05}$ & False\\
146 & 0.3913 $\pm$ 0.0001 & 42.91 $^{+0.61}_{-1.45}$ & 7.75 $^{+0.96}_{-0.87}$ & N.A. & 9.23 $^{+0.08}_{-0.09}$ & True\\
147 & 0.3890 $\pm$ 0.0022 & 8.37 $^{+1.87}_{-6.73}$ & 2.20 $^{+6.26}_{-1.38}$ & 8.57 $^{+0.75}_{-0.28}$ & 8.67 $^{+0.64}_{-0.45}$ & False\\
219 & 0.3984 $\pm$ 0.0004 & 40.25 $^{+9.93}_{-8.35}$ & 19.86 $^{+6.30}_{-7.32}$ & N.A. & 10.57 $^{+0.05}_{-0.10}$ & True\\
227 & 0.4005 $\pm$ 0.0002 & 11.56 $^{+1.87}_{-1.61}$ & 6.09 $^{+1.20}_{-1.45}$ & N.A. & 9.42 $^{+0.10}_{-0.06}$ & True\\
263 & 0.3995 $\pm$ 0.0002 & 54.41 $^{+5.41}_{-5.56}$ & 13.70 $^{+3.32}_{-3.27}$ & N.A. & 9.68 $^{+0.04}_{-0.05}$ & True\\
265 & 0.4006 $\pm$ 0.0002 & 26.58 $^{+2.73}_{-2.75}$ & 6.50 $^{+1.73}_{-1.86}$ & N.A. & 9.43 $^{+0.08}_{-0.14}$ & True\\
282 & 0.3802 $\pm$ 0.0001 & 12.72 $^{+0.78}_{-0.72}$ & 0.76 $^{+0.61}_{-0.26}$ & 8.20 $^{+0.10}_{-0.11}$ & 8.92 $^{+0.31}_{-0.31}$ & False\\
288 & 0.3911 $\pm$ 0.0001 & 23.85 $^{+0.01}_{-1.32}$ & 7.78 $^{+1.02}_{-0.67}$ & N.A. & 10.37 $^{+0.07}_{-0.08}$ & True\\
290 & 0.3797 $\pm$ 0.0004 & 8.81 $^{+1.46}_{-1.40}$ & 0.44 $^{+0.94}_{-0.44}$ & N.A. & 8.61 N.A. & True\\
291 & 0.3891 $\pm$ 0.0001 & 15.04 $^{+0.80}_{-0.67}$ & 6.23 $^{+0.69}_{-0.64}$ & N.A. & 9.29 $^{+0.23}_{-0.28}$ & True\\
308 & 0.3998 $\pm$ 0.0002 & 39.78 $^{+3.96}_{-3.65}$ & 3.87 $^{+1.84}_{-1.96}$ & 8.32 $^{+0.12}_{-0.23}$ & N.A. & False\\
317 & 0.4013 $\pm$ 0.0003 & 7.51 $^{+1.44}_{-1.77}$ & 3.49 $^{+1.26}_{-1.25}$ & 8.71 $^{+0.15}_{-0.17}$ & 9.33 $^{+0.27}_{-0.27}$ & False\\
336 & 0.3795 $\pm$ 0.0001 & 12.70 $^{+0.76}_{-0.89}$ & 1.93 $^{+0.66}_{-0.75}$ & 8.43 $^{+0.09}_{-0.13}$ & 9.89 $^{+0.12}_{-0.19}$ & False\\
338 & 0.3915 $\pm$ 0.0004 & 34.63 $^{+6.85}_{-6.71}$ & 11.02 $^{+5.33}_{-5.04}$ & 8.62 $^{+0.16}_{-0.27}$ & 10.62 $^{+0.03}_{-0.03}$ & False\\
339 & 0.3901 $\pm$ 0.0002 & 11.08 $^{+1.16}_{-1.48}$ & 0.29 $^{+1.03}_{-0.29}$ & 8.00 $^{+0.21}_{-0.36}$ & 9.72 $^{+0.04}_{-0.06}$ & False\\
341 & 0.3938 $\pm$ 0.0003 & 36.31 $^{+5.60}_{-5.53}$ & 7.75 $^{+3.75}_{-3.67}$ & N.A. & 10.40 $^{+0.06}_{-0.08}$ & True\\
343 & 0.3999 $\pm$ 0.0002 & 22.04 $^{+2.97}_{-2.66}$ & 7.86 $^{+2.15}_{-1.92}$ & N.A. & 10.40 $^{+0.05}_{-0.05}$ & True\\
345 & 0.3955 $\pm$ 0.0002 & 15.61 $^{+2.63}_{-2.72}$ & 5.07 $^{+1.98}_{-1.91}$ & 8.62 $^{+0.13}_{-0.17}$ & 10.35 $^{+0.04}_{-0.04}$ & False\\
353 & 0.3949 $\pm$ 0.0002 & 8.15 $^{+1.28}_{-1.23}$ & 1.79 $^{+0.84}_{-0.83}$ & 8.52 $^{+0.14}_{-0.25}$ & 9.23 $^{+0.23}_{-0.28}$ & False\\
366 & 0.3996 $\pm$ 0.0001 & 62.84 $^{+5.62}_{-5.36}$ & 11.90 $^{+2.95}_{-3.51}$ & 8.49 $^{+0.07}_{-0.09}$ & 9.74 $^{+0.05}_{-0.06}$ & False\\
384 & 0.3930 $\pm$ 0.0002 & 12.78 $^{+0.92}_{-0.96}$ & 2.21 $^{+0.70}_{-0.75}$ & N.A. & N.A. & True\\
405 & 0.3940 $\pm$ 0.0003 & 111.77 $^{+19.97}_{-18.56}$ & 74.77 $^{+14.71}_{-15.12}$ & 8.80 $^{+0.08}_{-0.10}$ & 10.96 $^{+0.03}_{-0.03}$ & False\\
422 & 0.3935 $\pm$ 0.0002 & 68.49 $^{+7.88}_{-8.92}$ & 25.64 $^{+5.67}_{-5.02}$ & 8.66 $^{+0.07}_{-0.08}$ & 10.02 $^{+0.04}_{-0.03}$ & False\\
424 & 0.3908 $\pm$ 0.0002 & 68.03 $^{+7.25}_{-7.15}$ & 12.45 $^{+3.92}_{-4.59}$ & 8.48 $^{+0.10}_{-0.14}$ & 10.22 $^{+0.04}_{-0.03}$ & False\\
433 & 0.4005 $\pm$ 0.0002 & 16.44 $^{+1.87}_{-1.65}$ & 1.03 $^{+0.73}_{-0.89}$ & 8.21 $^{+0.15}_{-0.40}$ & 8.87 $^{+0.42}_{-0.70}$ & False\\
443 & 0.3968 $\pm$ 0.0003 & 10.26 $^{+1.72}_{-1.84}$ & 1.72 $^{+1.09}_{-1.15}$ & 8.46 $^{+0.17}_{-0.48}$ & N.A. & False\\
451 & 0.3911 $\pm$ 0.0002 & 29.11 $^{+4.03}_{-3.86}$ & 3.02 $^{+2.03}_{-2.07}$ & 8.34 $^{+0.17}_{-0.39}$ & 10.09 $^{+0.08}_{-0.09}$ & False\\
456 & 0.3967 $\pm$ 0.0006 & 55.76 $^{+11.60}_{-14.53}$ & 29.93 $^{+12.46}_{-7.86}$ & N.A. & 10.70 $^{+0.04}_{-0.04}$ & True\\
457 & 0.3922 $\pm$ 0.0001 & 13.10 $^{+0.95}_{-0.92}$ & 11.13 $^{+0.92}_{-0.96}$ & N.A. & 10.60 $^{+0.03}_{-0.04}$ & True\\
485 & 0.3935 $\pm$ 0.0001 & 12.62 $^{+0.78}_{-0.74}$ & 2.98 $^{+0.75}_{-0.64}$ & 8.54 $^{+0.06}_{-0.07}$ & 9.08 $^{+0.29}_{-0.25}$ & False\\
501 & 0.3791 $\pm$ 0.0002 & 8.84 $^{+0.88}_{-0.80}$ & 3.69 $^{+0.67}_{-0.78}$ & 8.68 $^{+0.07}_{-0.07}$ & 10.24 $^{+0.12}_{-0.16}$ & False\\
560 & 0.3947 $\pm$ 0.0002 & 30.94 $^{+2.85}_{-3.15}$ & 4.50 $^{+1.76}_{-2.05}$ & 8.42 $^{+0.13}_{-0.21}$ & 9.59 $^{+0.05}_{-0.05}$ & False\\
574 & 0.3931 $\pm$ 0.0002 & 11.39 $^{+1.17}_{-1.14}$ & 1.50 $^{+0.71}_{-0.80}$ & 8.40 $^{+0.14}_{-0.26}$ & 9.20 $^{+0.50}_{-0.30}$ & False\\
582 & 0.3959 $\pm$ 0.0003 & 10.66 $^{+1.59}_{-1.53}$ & 2.68 $^{+1.07}_{-1.06}$ & 8.56 $^{+0.14}_{-0.20}$ & N.A. & False\\
612 & 0.3956 $\pm$ 0.0004 & 18.24 $^{+3.25}_{-2.52}$ & 4.39 $^{+1.58}_{-2.40}$ & N.A. & 10.00 $^{+0.05}_{-0.06}$ & True\\
651 & 0.3945 $\pm$ 0.0001 & 27.06 $^{+1.00}_{-1.02}$ & 8.48 $^{+0.87}_{-0.78}$ & N.A. & 10.42 $^{+0.06}_{-0.07}$ & True\\
657 & 0.3896 $\pm$ 0.0001 & 20.63 $^{+0.75}_{-0.79}$ & 6.99 $^{+0.71}_{-0.83}$ & 8.63 $^{+0.03}_{-0.04}$ & 9.83 $^{+0.08}_{-0.10}$ & False\\
658 & 0.3950 $\pm$ 0.0002 & 11.21 $^{+1.14}_{-1.22}$ & 0.36 $^{+0.61}_{-0.36}$ & N.A. & 9.53 $^{+0.12}_{-0.12}$ & True\\
664 & 0.3949 $\pm$ 0.0001 & 28.42 $^{+1.25}_{-1.13}$ & 8.90 $^{+0.99}_{-0.85}$ & N.A. & 10.33 $^{+0.07}_{-0.11}$ & True\\
675 & 0.3943 $\pm$ 0.0002 & 70.70 $^{+8.33}_{-8.22}$ & 25.60 $^{+5.26}_{-6.03}$ & 8.65 $^{+0.07}_{-0.08}$ & 10.38 $^{+0.04}_{-0.03}$ & False\\
677 & 0.3911 $\pm$ 0.0006 & 60.72 $^{+16.14}_{-16.58}$ & 21.82 $^{+12.43}_{-12.69}$ & N.A. & 10.75 $^{+0.08}_{-0.05}$ & True\\
700 & 0.3819 $\pm$ 0.0002 & 15.86 $^{+1.01}_{-1.08}$ & 3.30 $^{+0.89}_{-0.85}$ & N.A. & N.A. & True\\
708 & 0.3965 $\pm$ 0.0001 & 12.40 $^{+0.80}_{-0.61}$ & 1.72 $^{+0.64}_{-0.78}$ & 8.41 $^{+0.09}_{-0.15}$ & 9.72 $^{+0.17}_{-0.18}$ & False\\
714 & 0.3962 $\pm$ 0.0002 & 26.64 $^{+3.25}_{-3.29}$ & 8.08 $^{+2.20}_{-2.16}$ & 8.60 $^{+0.09}_{-0.10}$ & 10.00 $^{+0.08}_{-0.10}$ & False\\
728 & 0.3938 $\pm$ 0.0003 & 17.70 $^{+2.52}_{-2.45}$ & 7.60 $^{+1.68}_{-1.80}$ & N.A. & 9.87 $^{+0.10}_{-0.10}$ & True\\
758 & 0.3970 $\pm$ 0.0004 & 16.90 $^{+3.27}_{-3.24}$ & 9.79 $^{+2.35}_{-2.64}$ & N.A. & 10.42 $^{+0.08}_{-0.08}$ & True\\
773 & 0.3952 $\pm$ 0.0002 & 15.95 $^{+1.91}_{-1.87}$ & 0.74 $^{+0.98}_{-0.56}$ & 8.14 $^{+0.15}_{-0.32}$ & 8.94 $^{+0.51}_{-0.38}$ & False\\
775 & 0.3922 $\pm$ 0.0005 & 5.92 $^{+1.10}_{-0.95}$ & 0.96 $^{+0.58}_{-0.67}$ & 8.45 $^{+0.15}_{-0.47}$ & 8.86 $^{+0.29}_{-0.44}$ & False\\
777 & 0.3806 $\pm$ 0.0002 & 43.72 $^{+4.53}_{-4.32}$ & 10.48 $^{+2.39}_{-2.84}$ & 8.55 $^{+0.08}_{-0.10}$ & N.A. & False\\
783 & 0.4010 $\pm$ 0.0009 & 6.22 $^{+2.44}_{-2.59}$ & 1.41 $^{+1.64}_{-1.41}$ & 8.53 $^{+0.27}_{-0.79}$ & N.A. & False\\
811 & 0.4042 $\pm$ 0.0005 & 7.33 $^{+1.13}_{-1.23}$ & 0.71 $^{+0.72}_{-0.71}$ & N.A. & 10.45 $^{+0.08}_{-0.10}$ & True\\
826 & 0.3964 $\pm$ 0.0003 & 39.67 $^{+4.01}_{-4.27}$ & 9.71 $^{+2.72}_{-2.56}$ & N.A. & 9.57 $^{+0.24}_{-0.05}$ & True\\
847 & 0.3962 $\pm$ 0.0001 & 31.59 $^{+0.55}_{-0.56}$ & 2.41 $^{+0.56}_{-0.74}$ & 8.26 $^{+0.07}_{-0.09}$ & 9.41 $^{+0.07}_{-0.09}$ & False\\
850 & 0.3922 $\pm$ 0.0002 & 89.29 $^{+8.62}_{-9.25}$ & 24.36 $^{+5.94}_{-5.71}$ & N.A. & 10.48 $^{+0.05}_{-0.05}$ & True\\
869 & 0.3949 $\pm$ 0.0004 & 17.21 $^{+3.35}_{-2.34}$ & 5.64 $^{+1.72}_{-2.30}$ & N.A. & 9.95 $^{+0.11}_{-0.10}$ & True\\
872 & 0.3921 $\pm$ 0.0004 & 7.13 $^{+1.86}_{-1.87}$ & 4.34 $^{+1.73}_{-1.54}$ & 8.78 $^{+0.16}_{-0.18}$ & 10.41 $^{+0.07}_{-0.08}$ & False\\
875 & 0.3898 $\pm$ 0.0002 & 33.48 $^{+3.32}_{-3.53}$ & 3.54 $^{+1.87}_{-2.10}$ & 8.34 $^{+0.13}_{-0.29}$ & 9.89 $^{+0.08}_{-0.12}$ & False\\
882 & 0.4045 $\pm$ 0.0012 & 6.01 $^{+3.04}_{-3.39}$ & 2.54 $^{+2.58}_{-2.30}$ & 8.69 $^{+0.35}_{-0.74}$ & 10.24 $^{+0.15}_{-0.07}$ & False\\
889 & 0.3959 $\pm$ 0.0003 & 5.79 $^{+1.36}_{-1.34}$ & 2.79 $^{+1.19}_{-1.15}$ & N.A. & 10.27 $^{+0.12}_{-0.18}$ & True\\
919 & 0.3949 $\pm$ 0.0001 & 15.90 $^{+0.80}_{-0.82}$ & 1.73 $^{+0.64}_{-0.66}$ & 8.35 $^{+0.10}_{-0.15}$ & 10.10 $^{+0.09}_{-0.10}$ & False\\
927 & 0.3971 $\pm$ 0.0002 & 40.35 $^{+4.60}_{-4.48}$ & 11.91 $^{+2.78}_{-3.27}$ & N.A. & 10.08 $^{+0.05}_{-0.08}$ & True\\
929 & 0.3806 $\pm$ 0.0004 & 49.44 $^{+9.89}_{-8.84}$ & 20.88 $^{+6.70}_{-7.35}$ & 8.69 $^{+0.13}_{-0.18}$ & N.A. & False\\
938 & 0.3991 $\pm$ 0.0004 & 8.50 $^{+1.14}_{-1.21}$ & 1.73 $^{+0.62}_{-0.71}$ & 8.51 $^{+0.13}_{-0.20}$ & N.A. & False\\
966 & 0.3966 $\pm$ 0.0002 & 52.99 $^{+6.31}_{-5.55}$ & 24.49 $^{+3.54}_{-4.54}$ & 8.71 $^{+0.06}_{-0.07}$ & 10.29 $^{+0.05}_{-0.07}$ & False\\
977 & 0.3915 $\pm$ 0.0003 & 8.55 $^{+1.55}_{-1.61}$ & 3.51 $^{+1.24}_{-1.32}$ & N.A. & 9.84 $^{+0.05}_{-0.06}$ & True\\
998 & 0.3914 $\pm$ 0.0001 & 55.39 $^{+1.27}_{-1.20}$ & 10.26 $^{+1.14}_{-1.02}$ & N.A. & 10.22 $^{+0.08}_{-0.08}$ & True\\
1015 & 0.3936 $\pm$ 0.0001 & 83.94 $^{+0.01}_{-1.48}$ & 16.44 $^{+0.84}_{-0.64}$ & 8.50 $^{+0.02}_{-0.01}$ & 10.02 $^{+0.08}_{-0.07}$ & False\\
1039 & 0.3956 $\pm$ 0.0001 & 50.48 $^{+1.37}_{-1.32}$ & 23.94 $^{+0.30}_{-1.38}$ & N.A. & 10.82 $^{+0.06}_{-0.06}$ & True\\
1057 & 0.3933 $\pm$ 0.0001 & 20.64 $^{+0.82}_{-0.72}$ & 5.54 $^{+0.63}_{-0.60}$ & 8.57 $^{+0.03}_{-0.03}$ & 9.96 $^{+0.07}_{-0.11}$ & False\\
1097 & 0.3946 $\pm$ 0.0001 & 12.84 $^{+0.50}_{-0.56}$ & 2.40 $^{+0.48}_{-0.37}$ & 8.48 $^{+0.05}_{-0.05}$ & 9.44 $^{+0.32}_{-0.27}$ & False\\
1103 & 0.3959 $\pm$ 0.0001 & 11.65 $^{+0.72}_{-0.71}$ & 0.83 $^{+0.50}_{-0.50}$ & 8.25 $^{+0.11}_{-0.22}$ & 8.60 $^{+0.71}_{-0.39}$ & False\\
1125 & 0.3917 $\pm$ 0.0001 & 9.36 $^{+0.76}_{-0.69}$ & 2.61 $^{+0.58}_{-0.55}$ & N.A. & 9.16 $^{+0.15}_{-0.17}$ & True\\
1130 & 0.3918 $\pm$ 0.0001 & 10.76 $^{+0.60}_{-0.54}$ & 3.32 $^{+0.44}_{-0.38}$ & 8.61 $^{+0.04}_{-0.04}$ & 9.47 $^{+0.11}_{-0.12}$ & False\\
1165 & 0.3950 $\pm$ 0.0001 & 27.32 $^{+1.14}_{-1.13}$ & 8.58 $^{+0.91}_{-0.95}$ & 8.61 $^{+0.04}_{-0.04}$ & 10.56 $^{+0.04}_{-0.04}$ & False\\
1173 & 0.3901 $\pm$ 0.0001 & 39.05 $^{+0.69}_{-1.09}$ & 1.65 $^{+1.02}_{-0.54}$ & 8.12 $^{+0.08}_{-0.10}$ & 9.02 $^{+0.54}_{-0.19}$ & False\\
1204 & 0.3950 $\pm$ 0.0001 & 13.21 $^{+0.72}_{-0.81}$ & 1.73 $^{+0.65}_{-0.71}$ & 8.40 $^{+0.10}_{-0.15}$ & 8.61 $^{+0.52}_{-0.36}$ & False\\
1219 & 0.3822 $\pm$ 0.0001 & 33.56 $^{+0.01}_{-1.02}$ & 9.30 $^{+0.92}_{-0.96}$ & N.A. & 10.40 $^{+0.04}_{-0.04}$ & True\\
1220 & 0.3995 $\pm$ 0.0001 & 27.69 $^{+1.21}_{-1.13}$ & 10.76 $^{+0.80}_{-0.91}$ & 8.67 $^{+0.03}_{-0.03}$ & 9.92 $^{+0.07}_{-0.09}$ & False\\
41000 & 0.3809 $\pm$ 0.0002 & 19.24 $^{+1.53}_{-1.30}$ & 2.06 $^{+1.00}_{-1.12}$ & N.A. & N.A. & True\\

\hline

    \label{Tablagorda}
\end{longtable}
}

\end{appendix}

\end{document}